\documentclass{jfm}
\usepackage{graphicx}
\usepackage{newtxtext}
\usepackage{newtxmath}
\usepackage{natbib}
\usepackage{hyperref}
\hypersetup{
    colorlinks = true,
    urlcolor   = blue,
    citecolor  = black,
}

\newcommand{\RomanNumeralCaps}[1]
\title{Probabilistic thresholds of turbulence decay \\ in transitional shear flows}

\author{Daniel Morón\aff{1}
  \corresp{\email{daniel.moron@zarm.uni-bremen.de},}
  Alberto Vela-Mart\'in \aff{2}
 \and Marc Avila \aff{1,3}}

\affiliation{\aff{1}University of Bremen, Center of Applied Space Technology and Microgravity (ZARM), Am Fallturm 2, 28359 Bremen, Germany.
\aff{2} Universidad Carlos III de Madrid, Department of Aerospace Engineering, 28911 Leganés, Spain.
\aff{3} University of Bremen, MAPEX Center for Materials and Processes, Am Biologischen Garten 2, 28359 Bremen, Germany.}

\begin{document}
\maketitle
%%%%%%%%%%%%%%%%%%%%%%%%%%%%%%%%%%%%%%%%%%%%%%%%%%%%%%%%%%%%%%%%%
%                           ABSTRACT                            %
%%%%%%%%%%%%%%%%%%%%%%%%%%%%%%%%%%%%%%%%%%%%%%%%%%%%%%%%%%%%%%%%%
\begin{abstract}
Linearly stable shear flows first transition to turbulence in the form of localised patches. At low Reynolds numbers, these turbulent patches tend to suddenly decay, following a memoryless process typical of rare events. {How far in advance their decay can be forecasted is still unknown.} We perform massive ensembles of simulations of pipe flow and a reduced order model of shear flows \citep{moehlis2004low} and determine the first moment in time at which decay becomes fully predictable, subject to a given magnitude of the uncertainty on the flow state. By extensively sampling the chaotic sets, we find that, as one goes back in time from the point of inevitable decay, predictability degrades at greatly varying speeds. However, a well-defined (average) rate of predictability loss can be computed. This rate is independent of the uncertainty and also of the type of rare event, i.e. it applies to decay and to other extreme events. We leverage our databases to define thresholds that approximately separate phase-space regions of distinct decay predictability. Our study has implications for the development of predictive models, in particular it sets their theoretical limits. It also opens avenues to study the causes of extreme events in turbulent flows: a state which is predictable to produce an extreme event, it is causal to it from a probabilistic perspective.
\end{abstract}

%%%%%%%%%%%%%%%%%%%%%%%%%%%%%%%%%%%%%%%%%%%%%%%%%%%%%%%%%%%%%%%%%
%                        INTRODUCTION                           %
%%%%%%%%%%%%%%%%%%%%%%%%%%%%%%%%%%%%%%%%%%%%%%%%%%%%%%%%%%%%%%%%%
\section{Introduction}
\begin{figure}
\centering
\includegraphics[width=\textwidth, trim=0mm 0mm 0mm 0mm, clip=true]{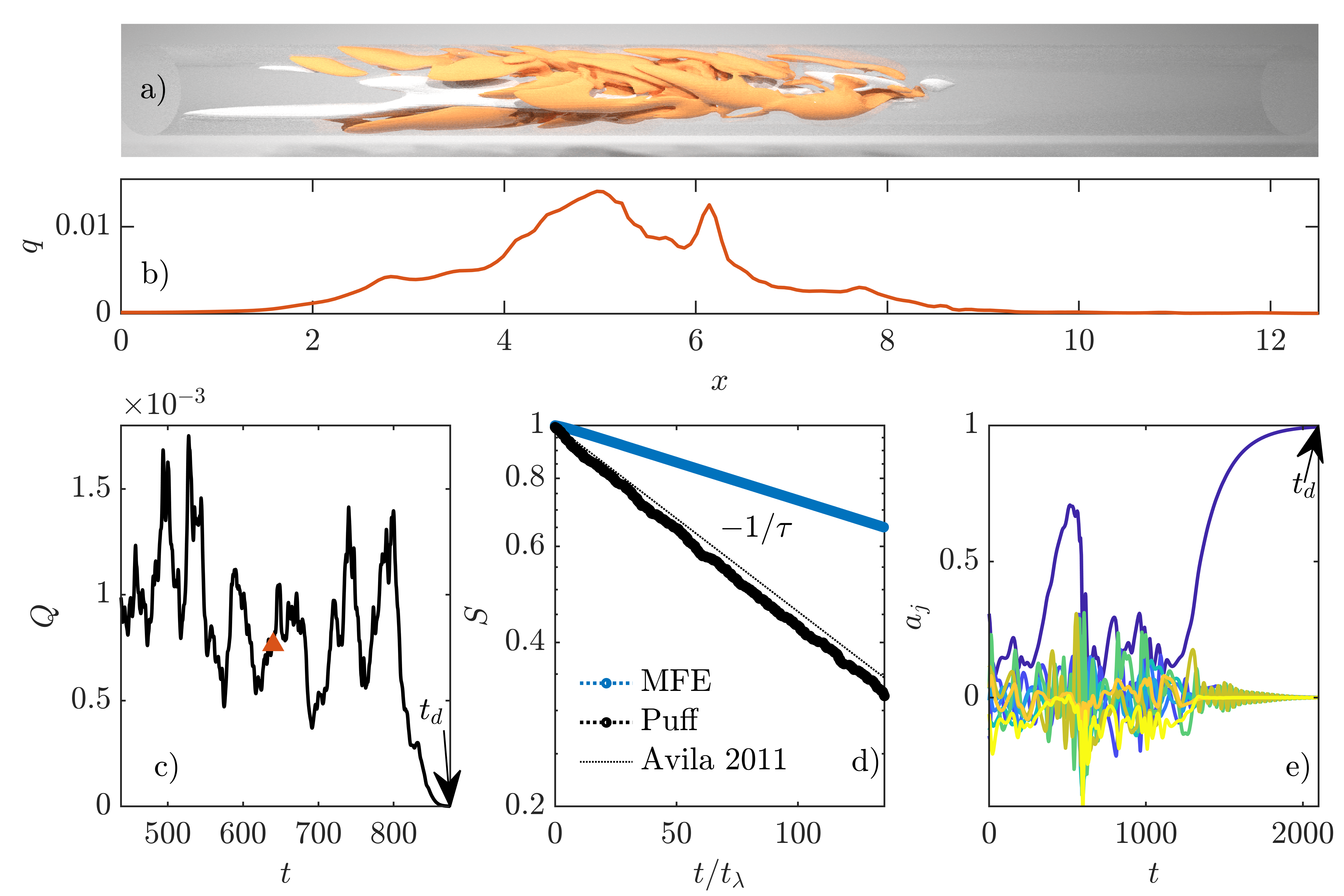}
\caption{Decaying events in transitional shear flows. a) snapshot of a turbulent puff at $Re=1850$. Grey denotes low axial velocity streaks $u_{x}' \approx -0.4 $. Red denotes regions where $u_{r}^{2}+u_{\theta}^{2} \geq 0.02$. b) cross-section averaged, cross flow kinetic energy $q=\left \langle u_{r}^{2}+u_{\theta}^{2} \right \rangle_{r,\theta}$, of the puff in a). c) volume averaged cross-section kinetic energy, eq.~\eqref{eq:CH2:Q}, of a decaying puff. The marker denotes the instant of time shown in the snapshot and $t_{d}$ stands for the time at which we detect decay, eq.~\eqref{eq:CH2:heuristic_pdecay}. d) survival probability of the MFE model ($Re=400$) and puffs ($Re=1850$). Time is normalized by the corresponding Lyapunov time. The dotted line corresponds to the value of the exponential distribution proposed by \cite{avila}, that fits the experimental data of \cite{Hof2008}. Here $S \approx \exp\left(-t/\tau \right)$, where $\tau = \exp \left( \exp \left(  5.56~10^{-3} \cdot Re-8.5 \right) \right)$. {e}) decaying trajectory of the MFE model \citep{moehlis2004low}. Lighter color means higher $j$.}
\label{fig:fig1}
\end{figure}

Over 140 years {after} Reynolds' experiment \citep{Reynolds1883a}, the transition to turbulence of Newtonian fluids flowing in a rigid smooth pipe continues to puzzle researchers, (e.g. the recent review by \cite{AvilaAnnRev}). The problem depends on the Reynolds number $Re=\frac{UD}{\nu}$, where $U$ is the bulk velocity, $D$ the pipe diameter and $\nu$ the kinematic viscosity of the fluid; and on the amplitude and shape of the disturbances to the laminar flow. Specifically, pipe flow is linearly stable for {at least} $Re \leq 10^{7}$ \citep{Meseguer2003a} but still, when sufficiently perturbed, it transitions to turbulence at $Re \approx 2000$. It first does so in the form of localized turbulent {patches} known as turbulent puffs \citep{wygnanski1973transition}. We show a snapshot of a turbulent puff in figure~\ref{fig:fig1}a. Once triggered, the puff dynamics become independent of the disturbance that created it. Turbulent puffs can either elongate and split thereby increasing the turbulent fraction; or suddenly decay resulting in laminar flow, fig.~\ref{fig:fig1}c. Decay events are more frequent than split events at $Re \lesssim 2040$ \citep{avila}, and obey a memoryless process: the statistics of decay events follow an exponential distribution \citep{eckhardt2004} with a mean lifetime $\tau$ that scales super-exponentially with the Reynolds number \citep{Hof2008,avila2010transient}. Lifetime statistics at $Re=1850$ are shown in figure~\ref{fig:fig1}d. \cite{goldenfeld2010extreme} theorized that the super-exponential scaling of the mean lifetime of decay events, can be explained {through} extreme value theory. \cite{Alexakis}, derived the super-exponential scaling by assuming that relaminarisation is linked to extremely low axial vorticity events throughout the structure of the puff. However, their study raises the following questions. Are extremely low values of the axial vorticity a cause or a consequence of decay events? Is their proposed threshold for decay actually a sufficient, or even a necessary, condition for decay?

Memoryless decay events are not limited to pipe flow, and are also found in other transitional shear flows such as Taylor--Couette \citep{borrero2010transient} and Couette \citep{bottin,shi2013} flows. {In channel flows, turbulent bands also decay following a memoryless process in short spanwise \citep{shimizu2019exponential} and tilted \citep{tuckerman2020patterns} domains. In long domains the decay statistics are not memoryless but depend on the history of the turbulent bands \citep{xu2022size}. However, when the length of the turbulent bands is fixed, decay events follow a memoryless process too \citep{wu2025transient}. This supports the idea that the behavior of transiently chaotic turbulent states is universal among all subcritical shear flows in the transitional regime \citep{rempel2010supertransient,linkmann2015sudden}.}

{Reduced order models of shear flows are also able to capture these decay events. \citet{willis2009turbulent} defined a reduced order model of pipe flow by truncating their Galerkin approximation to a single azimuthal Fourier mode. Their model retained memoryless puff decay. Another example,} is the reduced order model developed by \citet{moehlis2004low}, hereafter referred to as the MFE model. This model is {a nine-dimensional ODE obtained from} a truncated {Galerkin projection} of the flow between two parallel free-slip walls, driven with a wall-normal sinusoidal force \citep{waleffe1997self}. An example of a trajectory {of this model} is shown in figure~\ref{fig:fig1}e, {and lifetime statistics in figure~\ref{fig:fig1}d.} \cite{lellep2022interpreted} investigated the decay events of the model using an explainable artificial intelligence method. They successfully performed predictions at short times before decay, but observed a quick degradation of their predictions as they considered longer times before the event. 

{Generally} there are two main challenges in predicting relaminarisation. On the one hand, there {are many routes} to decay {in phase space} \citep{Chantry_Schneider_2014,budanur2019geometry}. On the other hand, turbulent flows are chaotic \citep{lorenz} {and} the uncertainty in the determination of a flow state can result in radically different decay times. Without a correct assessment of these two issues, the ability of predictive models {is severely limited and thresholds of decay are necessarily conservative.}

In this paper, we examine the threshold of decay from a new probabilistic perspective: by studying predictability. We propose that the threshold of decay is the first state of a trajectory that becomes fully predictable to decay, given a finite size uncertainty \citep{palmer2000}. Uncertainties, have many different origins (measurements, model, numerical method...) and are unavoidable. We argue that, fully characterizing the predictability of decay is a prerequisite to successfully finding the threshold and causes of decay. Predictability has historically been characterized with the Lyapunov exponents \citep{lorenz}. They measure the rate at which trajectories, initially separated by an infinitesimally small uncertainty $\delta$, separate exponentially in phase space. However, they fail to correctly measure predictability, when $\delta$ is not infinitesimally small \citep{boffetta2002predictability}, or in the case of systems with several time scales \citep{aurell1996}, like turbulence. They are also general measurements of a chaotic system, and not specific to a single event of interest. Another, more powerful, alternative is to directly study the evolution of the probability distribution function of possible future states of the system. The Liouville conservation equation models the evolution of probability distribution functions and has an analytical solution \citep{liouville1838note}. However, it is unfeasible to define this equation for high dimensional systems like turbulence. Recently \cite{jimenez2023}, integrated the probability density function of a reduced representation of a turbulent channel flow using the Perron-Frobenius operator. However, it is difficult to replicate this analysis for other turbulent flows without knowledge of appropiate reduced-order projections of the flow. 

% Paragraph about albertos work
\cite{Vela_Martín_Avila_2024} proposed an alternative measurement of predictability. Inspired by weather forecasting \citep{palmer1993ensemble}, they performed massive ensembles of simulations {of the two-dimensional Kolmogorov flow,} and used a metric derived from information theory, the Kullback-Leibler divergence \citep{kullback1951information}, to characterize the predictability of extreme {dissipation} events. In a proof of concept \citep{Moron_2024}, we used this metric to characterize predictability of decay events in the MFE model and pipe flow. In this paper, we extend our previous study and find thresholds of decay both in time and in phase space using {massive ensembles and the Kullback-Leibler divergence to asses the predictability of decay.} The rest of the paper is structured as follows. In section~\ref{sec:Methods} we describe the MFE and puffs in pipe flow in more detail. In section~\ref{sec:Pred} we describe the method we use to characterize predictability, and discuss results of predictability for the two decay events of interest. In sections \ref{sec:MFEresults} and \ref{sec:Puffresults} we analyze MFE and puff trajectories, according to their predictability and in section \ref{sec:Conclusions} we draw our main conclusions.

%%%%%%%%%%%%%%%%%%%%%%%%%%%%%%%%%%%%%%%%%%%%%%%%%%%%%%%%%%%%%%%%%
%                         METHODOLOGY                           %
%%%%%%%%%%%%%%%%%%%%%%%%%%%%%%%%%%%%%%%%%%%%%%%%%%%%%%%%%%%%%%%%%
\section{Methods} \label{sec:Methods}

\subsection{Puffs in pipe flow}\label{sec:Puffsmethods}
We consider the flow of a viscous Newtonian fluid with constant properties in a straight smooth rigid pipe of circular cross-section. The flow is incompressible and governed by the dimensionless Navier--Stokes equations
\begin{equation}
\frac{\partial\pmb{u}}{\partial t} + \left(\pmb{u} \cdot \nabla \right) \pmb{u} =
-\nabla p + \frac{1}{Re}\nabla^{2}\pmb{u}+f_{p}\left(t\right)\cdot\pmb{e}_{x}
\quad\text{and}\quad
\nabla\cdot\pmb{u} = 0
\text{.}
\label{eq:CH2:NSeq}
\end{equation}
Here, $\pmb{u}$ is the fluid velocity, $p$ the pressure, $\pmb{e}_{x}$ the unit vector in the axial (stream-wise) direction and $f_{p}\left(t\right)$ the pressure gradient that drives the flow; $f_{p}\left(t\right)$ is adjusted at each time step to enforce a constant bulk velocity. All variables are rendered dimensionless using the pipe diameter ($D$) and the bulk velocity ($U$). The equations are formulated in cylindrical coordinates $\left(r,\theta,x\right)$, with velocity field components $\left(u_{r},u_{\theta},u_{x}\right)$ in the radial, azimuthal and axial directions, respectively. Throughout this paper we fix $Re=1850$. We define
\begin{equation}
Q \left(t\right)= \left \langle u_{r}^{2} + u_{\theta}^{2} \right \rangle_{V} \text{,}
\label{eq:CH2:Q}
\end{equation}
as the volume-averaged kinetic energy of the cross-sectional velocity, and
\begin{equation}
U_{c}' \left(t\right)= \left \langle u_{HP}\left(r=0\right)-u_{x}\left(r=0\right)  \right \rangle_{V} \text{,}
\label{eq:CH2:Uc}
\end{equation}
as the volume-averaged deviation from the Hagen--Poiseuille $u_{HP}$ center line velocity. We choose these two variables as they have been shown to capture the dynamics of transitional pipe flow very well \citep{barkley2011simplifying,barkley2015rise}.

As a heuristic threshold, we {define} puff decay at $t=t_{d}$ when
\begin{equation}
   Q \left(t \geq t_{d}\right) \leq 10^{-7} \text{, for } Q \left(t < t_{d}\right) > 10^{-7} \text{.}
   \label{eq:CH2:heuristic_pdecay}
\end{equation}

We solve eq.~\eqref{eq:CH2:NSeq} numerically using {our} GPU-CUDA pseudo-spectral code \citep{Moron_2024}, {which is} publicly available in this \href{https://github.com/Mordered/nsPipe-GPU}{Github}. We perform DNS in a $L_{x}=50D$ long pipe and use $N_{r}=48$ radial points, $M_{\theta}=96$ azimuthal and $M_{x}=768$ axial physical points after de-aliasing. The maximum wall Reynolds number measured is $Re_{\tau} \approx 70$ which results in a grid spacing in wall units of $0.06 \lesssim \Delta r^{+} \lesssim 2.2$, $D \Delta \theta^{+}/2 \approx 4.5 $ and $\Delta x^{+} \approx 9$. The time step size is set to $\Delta t = 0.0025 $. The resolution is optimized to reduce computing cost, while giving an accurate estimate of the lifetime, {see} figure~\ref{fig:fig1}d. 

\subsection{A reduced order model of shear flows}\label{sec:MFE}
The MFE model \citep{moehlis2004low} has nine {dimensionless} time dependent {modes} $a_{j}$ governed by a system of nonlinear {ODEs} of the form
\begin{equation}
    \frac{\mathrm{d} a_{j}}{\mathrm{d} t} = L_{j} a_{j} + N_{j} \left(\pmb{a} \right) + c_{j} \text{,}
\end{equation}
where, $L_{j}$ and $N_{j} \left( \pmb{a} \right)$ are respectively linear and quadratic operators, $c_{j}$ model constants and $\pmb{a}=\left(a_{1},a_{2},...\right)^{T}$. {All the variables are normalized by the channel half height $h/2$ and the laminar velocity $U_{0}$ at the walls.  (see the full equations in appendix~\ref{ap:MFEeq}).}

Each variable $a_{j}$ {is} the amplitude of one Fourier mode $\pmb{v}_{j}$. The {dimensionless} velocity field is computed as
\begin{equation}
    \pmb{u}\left(\pmb{x},t \right)= \sum_{j=1}^{9} a_{j}\left( t \right)\pmb{v}_{j}\left(\pmb{x}\right)\text{.}
\end{equation}
The modes $\pmb{v}_{j}$ have a direct interpretation in the flow. The mode $j=1$ represents the mean velocity profile. When $a_{1} \equiv 1$ the flow is laminar, and turbulent if $a_{1}<1$. The mode $j=2$ represents velocity streaks; $j=3$ streamwise vortices; $j=4$ and $j=5$ two spanwise flows; $j=6$ and $j=7$ two wall-normal vortices; and $j=9$ represents a correction to the mean velocity profile. The mode $j=8$ does not have a clear interpretation, but {is} a three-dimensional mode that interacts with the others. 

The MFE model depends on the Reynolds number, defined as $Re=\frac{U_{0}h}{2 \nu}$, and the spanwise/streamwise domain sizes $L_{z}$ and $L_{x}$. We fix $Re=400$, $L_{x}=4 \pi $ and $L_{z}=2 \pi$. We integrate the MFE model forward in time using a low-storage explicit {fourth} order Runge--Kutta method, with a constant time step size $\Delta t=0.05 $.

We define the energy of the mean profile as
\begin{equation}
E_{1}= \left(1-a_{1}\right)^{2}\text{,}
\label{eq:CH2:E1}
\end{equation}
and the energy of the fluctuations as
\begin{equation}
    E_{j}=\sum_{j=2}^{9} a_{j}^{2}\text{.}
    \label{eq:CH2:Ej}
\end{equation}
The flow is laminar if $E_{1}=E_{j}=0$. We {define} turbulence decay at $t=t_{d}$ when:
\begin{equation}
    a_{1}\left(t \geq t_{d}\right) \geq 0.995 \text{, for } a_{1}\left(t < t_{d} \right)<0.995 \text{.}
    \label{eq:CH2:heuristic_MFEdecay}
\end{equation}

%%%%%%%%%%%%%%%%%%%%%%%%%%%%%%%%%%%%%%%%%%%%%%%%%%%%%%%%%%%%%%%%%
%                       PREDICTABILITY                          %
%%%%%%%%%%%%%%%%%%%%%%%%%%%%%%%%%%%%%%%%%%%%%%%%%%%%%%%%%%%%%%%%%
\section{Forecasting of turbulence decay}\label{sec:Pred}
\begin{figure}
\centering
\includegraphics[width=\textwidth, trim=0mm 0mm 0mm 0mm, clip=true]{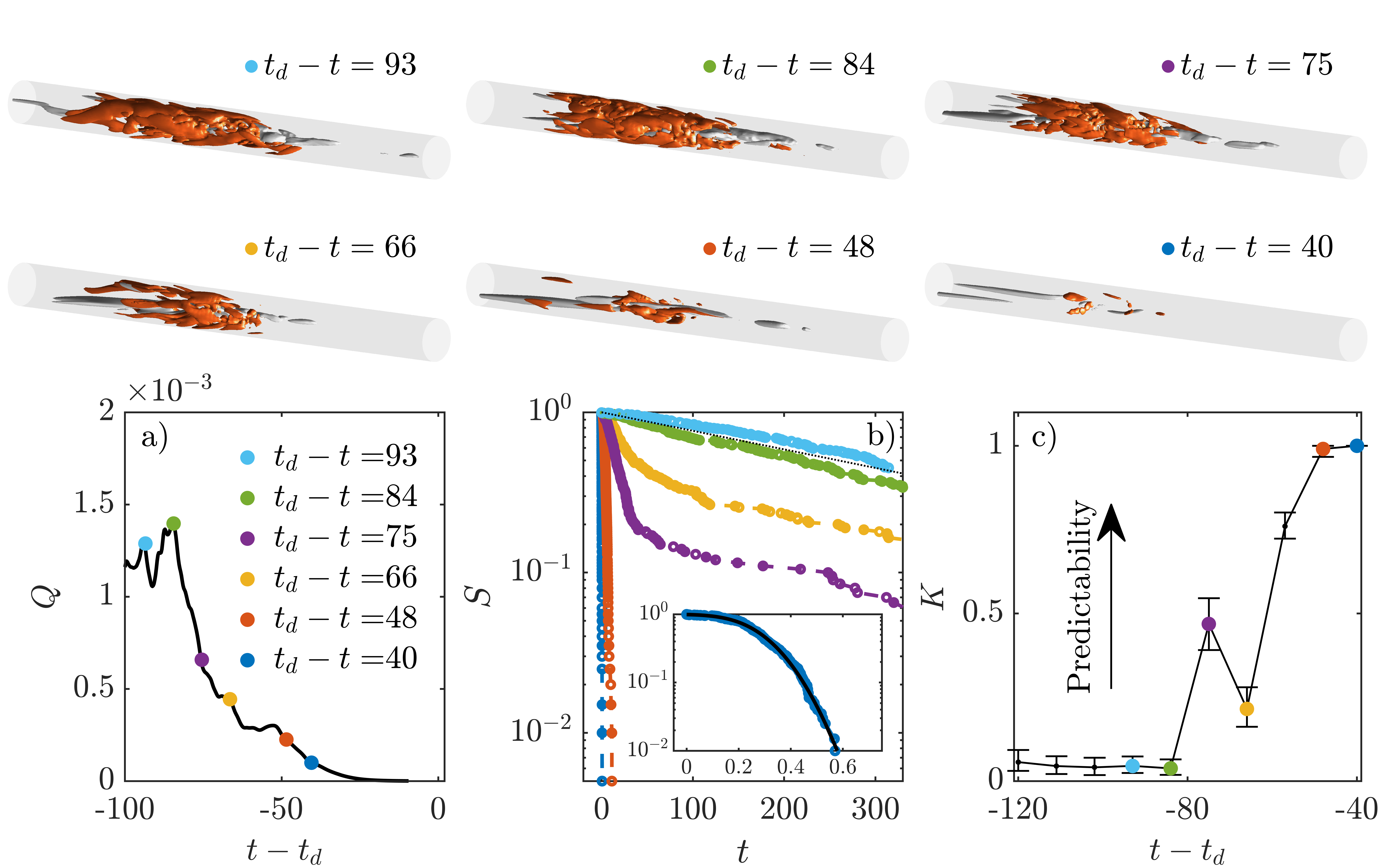}
\caption{Example of a puff decay event and {the method we use to asses its predictability}. At the top: snapshots of a turbulent puff as it decays. Grey denotes low axial velocity streaks $u_{x}' \approx -0.4$. Red denotes regions where $q=u_{r}^{2}+u_{\theta}^{2} \geq 0.02 $. The time before decay $t-t_{d}$ is indicated on top of each snapshot. a) zoom in to the puff decay trajectory shown in figure~\ref{fig:fig1}c. The snapshots at the top are indicated as colored markers in the plot. b) survival probability of the ensembles initialized about the instantaneous puffs shown in the snapshots. The color of the distributions correspond to the color of the markers in the snapshots and in the trajectory. With a dotted line, the exponential distribution, $S = 1-P_{q} \approx \exp\left(-t/\tau \right)$. The insect is a zoom in for the ensemble at $t-t_{d}=-40$. The black line is the cumulative function of a Gaussian distribution, $f=0.5-0.5 \text{erf}\left[b \left( t - c \right) \right]$, fitted to the data. Here $b$ and $c$ are fitted parameters and $\text{erf}$ stands for the error function. c) Time-dependent $K$, eq.~\eqref{eq:CH3:KLD}, of this puff trajectory. A higher $K$ means more predictability to decay. At $K\rightarrow 0$ the corresponding puff is fully unpredictable. The error bars represent the uncertainty of computing $K$ from our finite sample. The colored markers correspond to the instantaneous states shown in the snapshots and indicated in a).}
\label{fig:fig2}
\end{figure}

\subsection{Massive ensembles of simulations}
We run $i=1,...,N_{b}$ MFE{/pipe} simulations each initialized with a different chaotic state{/puff}. These are our base trajectory simulations, and for each {of them} we save $N_{t}$ instantaneous states at times $t_{k}$ (every $5 $ time units for the MFE model and $8 $ for pipe flow). We run each base trajectory, $i$, until we observe decay according to our conservative heuristic thresholds, eq.~\eqref{eq:CH2:heuristic_MFEdecay} for the MFE model, and eq.~\eqref{eq:CH2:heuristic_pdecay} for pipe flow. We save the time at which this threshold is reached for each individual base trajectory $i$, and call it the time of decay $t_{d}$, see fig.~\ref{fig:fig1}c and e. {Note that $t_{d}$ is different for each base trajectory and it is always larger than the times at which we sample the trajectories $t_{d}>t_{k}$}. We discarded base trajectories if $t_{d}$ is too short, $t_{d}\leq 3000$ in the MFE model and $t_{d}\leq 250$ in pipe flow. 

{For each of the $N_{b} \times N_{t}$ instantaneous states, we launch a massive ensemble of simulations. In the case of puffs, a simulation in an ensemble is initialized as }
\begin{equation}
\pmb{u} \left(t_{k}-t_{d}\right)= \pmb{u}_{t_{k}} + \epsilon_{0} \pmb{\mathcal{N}}\left(0,1\right) \text{,}
\end{equation} 
where $\pmb{u}_{t_{k}}$ is the velocity field of {the} base trajectory $i$ at time $t_{k}$, $\pmb{\mathcal{N}}$ {a} Gaussian noise with zero mean and unit standard deviation {applied at each velocity component and at each point of the domain}; and $\epsilon_{0}$ the magnitude of the noise. 

For the MFE model each simulation in the ensemble is initialized as: 
\begin{equation} 
\pmb{a} \left(t_{k}-t_{d}\right) = \pmb{a}_{t_{k}} + \epsilon_{0} \pmb{\gamma} \text{,}
\end{equation} 
where the 9-dimensional Gaussian random vector $\pmb{\gamma}$ has a module equal to 1. Crucially $\epsilon_{0}$ is always sufficiently small, so two initial conditions of the same ensemble are, from a macroscopic point of view identical. In particular $\epsilon_{0} \leq 10^{-3} $ in the MFE model, and $\epsilon_{0} \leq 10^{-2}$ in pipe flow.

For each combination of flow parameters and $\epsilon_{0}$, we perform $N_\text{total}=N_{b} \times N_{t} \times N_{e}$ individual simulations, {where}: $N_{b}$ {is} the number of base trajectories, $N_{t}$ the number of sampled times and $N_{e}$ the members per ensemble (see table~\ref{tab:CH3:param}). {Unless stated otherwise we set $\epsilon_{0} = 10^{-4} $ in the MFE model, and $\epsilon_{0} = 10^{-2}$ in pipe flow.} 

\subsection{Measuring the predictability of decay with the Kullback-Leibler divergence}

{We study lifetime statistics of the ensemble members. For this purpose, we compute $\rho $ as the probability distribution function of the ensemble lifetime, $P$ the corresponding cumulative distribution function, and $S=1-P$ the survivor function \citep{lawless2011statistical}. We denote $\rho_{q}$ as the exponential lifetime distribution of puffs in pipe flow, i.e. computed with random initial conditions across phase space \citep{eckhardt2004} and with mean lifetime $\tau$, see fig.~\ref{fig:fig1}d. While $\rho_{q}$ is {a characteristic} of the dynamical system, $\rho$ depends on the particular ensemble of interest, {i.e. is the lifetime distribution conditional to being computed using a known instantaneous state plus a small noise as initial condition}. We compute statistics using histograms with equispaced $20$ bins. All the distributions ($\rho$, $P$ and $S$) are horizontally shifted so the first bin corresponds to the earliest decay event observed in each ensemble.}

\begin{table}
\centering
\caption{Number of simulations performed for each system of interest. $N_{b}$ is the number of base trajectories, $N_{t}$ the number of sampled times, $N_{e}$ the members per ensemble and $N_\text{total}=N_{b} \times N_{t} \times N_{e}$ the total number of individual simulations.}
\label{tab:CH3:param}
\begin{tabular}{lllll}
\hline
              & $N_{b}$    & $N_{t}$       & $N_{e}$ & $N_\text{total}$ \\ \hline
Puffs        & 100       & 10         & 200 & $2\cdot 10^{5}$ \\
MFE       & 14000        & 400       & 1000 & $5.6 \cdot 10^{9}$ \\   \hline
\end{tabular}
\end{table}

In figure~\ref{fig:fig2}b, we show {the survivor functions of different ensembles, each initiated using a different sampled state at different times $t_{k}$ of} a specific base trajectory $i=1$. Note how the distribution of puff decay tends to be clustered close to $t=0$, if the state used to initialize the ensemble is close to decay {$\left(t- t_{d} \gtrsim -40 \right)$. Here}, all trajectories decay {at approximately the same time as} the base trajectory {does, and} the distribution of lifetimes inherits the Gaussian distribution of the initial conditions. {For such puffs, decay is inevitable. As one uses states farther back in time as initial conditions, the resulting survivor functions gradually tend towards the expected exponential distribution. For the specific base trajectory shown in figure~\ref{fig:fig2}b, $S$ (conversely $\rho$) becomes indistinguishable from $S_{q}$ ($\rho_{q}$) at $t -t_{d} \lesssim -80$.}

We measure predictability as the difference between the conditional probability $\rho$, and the exponential $\rho_{q}$ one. {Specifically}, we use the Kullback--Leibler divergence \citep{kullback1951information} defined as:
\begin{equation}
K = \frac{1}{\text{K}_\text{max}} \sum \rho  \log \left( \frac{ \rho}{\rho_{q}}  \right)  \text{,}
\label{eq:CH3:KLD}
\end{equation}
{where $\text{K}_\text{max}$ is a parameter, that normalizes $K$ to $K \leq 1$.} $K$ is minimum when $\rho \equiv \rho_{q}$; then $K \rightarrow 0$. The bigger $K$ is, the more different the two distributions are, and therefore, the more predictable the members of the ensemble become. {Since we compute statistics, and therefore $K$, using bins, $K$ is maximum $\left( K \equiv 1 \right)$, when the probability distribution function $\rho$ is clustered in a single bin. In our case, this happens when all the members of the ensemble decay at nearly the same time as the base trajectory. In figure~\ref{fig:fig2}, this happes for the puff at $t- t_{d} \approx -40$. The parameter $\text{K}_\text{max}$ is computed for this case, and depends on the number of bins we use. As we show in appendix~\ref{app:robust} our method is robust to different noise shapes, noise magnitudes and ensemble sizes. Overall,} $K$ measures the information we gain by knowing the initial state of the system up to an uncertainty $\epsilon_{0}$, compared to the information we have by assuming a random initial state.

For each base trajectory, $i=1,...,N_{b}$ and sampling time $k=1,...,N_{t}$, we compute $K$ of the corresponding ensemble of simulations. We assume that $K$ is continuous in time and study its time evolution. {In what follows, we refer to $K$ as the predictability.}

\subsection{Temporal evolution of the predictability}
\begin{figure}
\centering
\includegraphics[width=\textwidth, trim=0mm 0mm 0mm 0mm, clip=true]{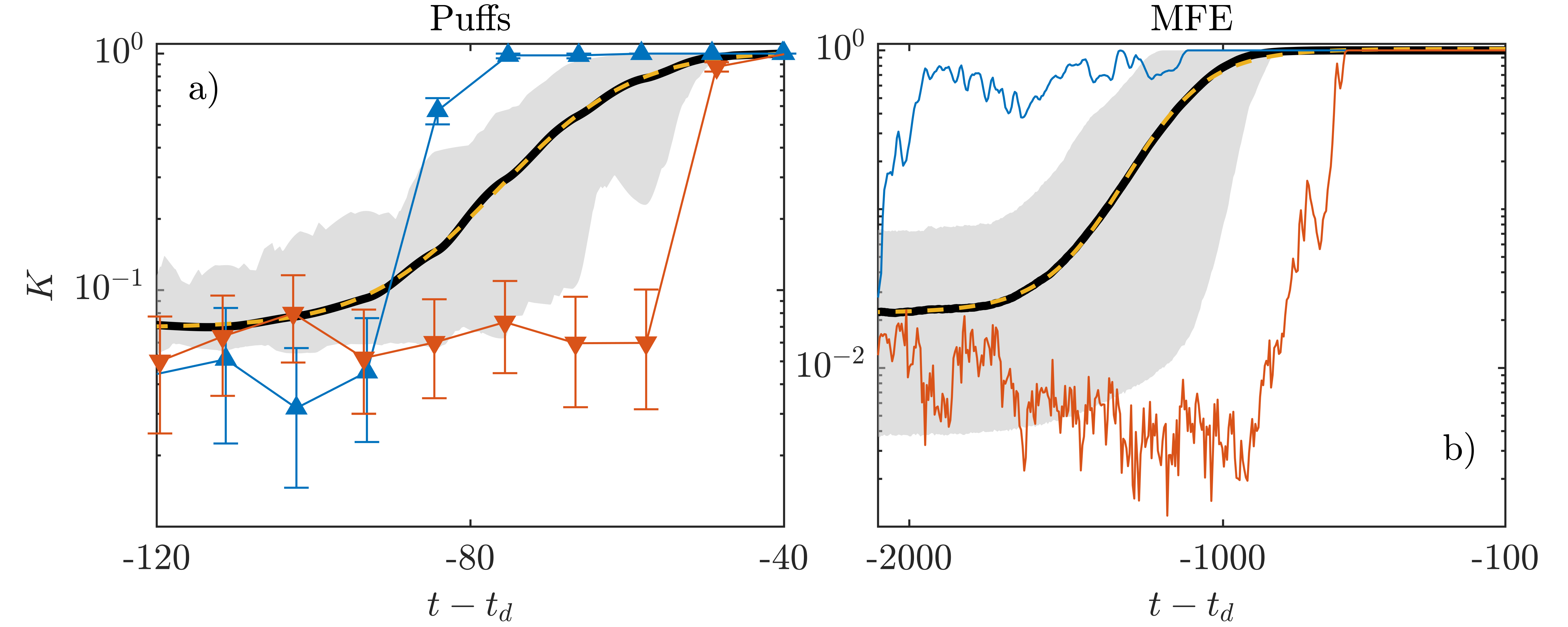}
\caption{Statistics of $K$ (as a measurement of predictability) with respect to time, for the two decay events of interest {(indicated in the title of the plot)}. In the two subplots, the solid black line denotes the mean predictability, and the dashed yellow line a fit of this mean predictability to equation~\eqref{eq:CH3:fitKLD}. The shaded region denotes the first and the last decile of the data. The red line corresponds to the case that was unpredictable for a longer time-span, and the blue line to the one that was predictable for the longest time. The errorbars stand for the uncertainty in the determination of $K$ after a bootstrapping analysis.}
\label{fig:CH3:mKLD}
\end{figure}

In figure~\ref{fig:fig2}c, we show the evolution of the {predictability for a specific} base trajectory. At sufficiently long times before the decay event, the lifetime distributions are similar to the exponential distribution, resulting in {a fully unpredictable decay $\left( K \rightarrow 0 \right)$}. As one uses as initial condition puffs that are closer to the decay event, $K$ increases. We report that this increase does not need to be monotonic, as the decaying trajectory visits regions of phase space with varying predictability. Finally $K$ saturates, implying the inevitability of decay for our chosen level of uncertainty. 

We repeat this analysis for all our MFE and puff base trajectories, see table~\ref{tab:CH3:param}, and compute statistics of $K$ with respect to time, see fig.~\ref{fig:CH3:mKLD}. To compare $K$ between different trajectories, here, and in the rest of the paper we define:
\begin{equation}
\Delta t_{d} = t - t_{d} \text{,}
\label{eq:deltatd}
\end{equation}
as our reference time frame. At the same time before a decay event, we observe $K$ that differ in orders of magnitude between trajectories, see fig.~\ref{fig:CH3:mKLD}. There is however a point in time where all the trajectories saturate, $\left( K \rightarrow 1 \right)$. In the case of the MFE this happens at $\Delta t_{d} \approx -800$ while for the puffs at $ \Delta t_{d} \approx -40$. This means that, for this level of uncertainty ($\epsilon_{0}$), this is the earliest time when one can perform a prediction to decay with almost perfect certainty, {for all base trajectories studied.}

The mean $K$, averaged among all the decaying trajectories, and represented as a thick black line in the figures, is approximated here as:
\begin{equation}
\left \langle K \right \rangle_{N_{b}} \approx A \tanh \left( \frac{\Delta t_{d}+t_{0}}{t_{c}} \right) - B \text{,}
\label{eq:CH3:fitKLD}
\end{equation}
where $A \approx B \approx 0.5$ are two dimensionless parameters, and $t_{0}$ and $t_{c}$ are physically relevant times. The formula fits the mean trends of $K$ reasonably well, as seen with the dashed yellow lines in fig.~\ref{fig:CH3:mKLD}. 

%We report that, for all the fits of $K$ to the formula~\eqref{eq:CH3:fitKLD} shown in this paper, $A \approx B \approx 0.5$. However $A$ and $B$ will never be equal to $0.5$ since the averaged $K$ never reaches a value of $K \equiv 0$. This is because both the MFE model and transitional pipe flow are transiently chaotic systems. This means that there are many regions of phase space that are predictable to decay \citep{budanur2019geometry}. Trajectories can pass close to these regions without subsequently decaying. But this passing will result in a slight increase in predictability. This is the reason why, in this particular problem, the lower bound of $K$ is not $K = 0$ but a different, albeit small, number.

\begin{table}
\centering
\caption{Time-scales of the two systems of interest.}
\label{tab:CH3:timsca}
\begin{tabular}{cccc}
\hline
Time scale   & Puff    & MFE        & Description  \\ \hline
$t_{\lambda}$& $2.95$  & $33.78$    & Inverse of mean Lyapunov exponent \\
$\tau$       & $285.95$& $1.05e4$   & Mean lifetime of the event \\
$t_{c}$      & $15.01$ & $264.08  $ & Mean time of predictability loss\\
$t_{0}$      & $66.35$ & $1314.1  $ & Mean predictability bias \\\hline
\end{tabular}
\end{table}
The time $t_{c}$, in equation~\eqref{eq:CH3:fitKLD}, is the mean time of predictability loss, see  table~\ref{tab:CH3:timsca}. It measures the time scale at which predictability degrades as one goes back in time with respect to the event of interest. We observe that $t_{c}$ is approximately one order of magnitude larger than the mean Lyapunov time. 

The time $t_{0}$ is the mean predictability bias. It represents the past time where, on average, predictability reaches half its maximum, and it is mostly affected by the magnitude of uncertainties $\epsilon_{0}$, {whereas $t_{c}$ is not}, see~\ref{sec:CH3:ep0}.

% THE MFE MODEL!!!!!!!!!!!!!!!!!!!!!!!!!!!!!!!!!!!!!!!!!!!!!!!!!
\section{Predictability of turbulence decay in the reduced order model}\label{sec:MFEresults}
In what follows we classify trajectories and instantaneous states of the MFE model according to their predictability {$\left(K\right)$. As shown in figure~\ref{fig:CH3:mKLD}, $K$ varies by several orders of magnitudes. Specifically, unpredictable cases have very low $K$ and therefore little weight when computing averages of predictability. To better represent the strong variations of $K$ and appropriately weight unpredictable cases, we consider the metric}
\begin{equation}
\kappa = \log \left( K \right) \text{.}
\label{eq:kappa}
\end{equation}
{in the subsequent analyses. The presented averages of $\kappa$ correspond to geometric averages of $K$ and better capture the natural variation of predictability in the flow.}

\subsection{Classification of model trajectories according to their predictability}\label{sec:MFEtraj}
{We here perform a conditional analysis of trajectories according to their predictability (low/high). For this purpose, we look for the first moment in time when} at least $N_{p}=1000$ trajectories have maximum $\kappa$. We find that this happens at $t_{p} \approx t_{d}-1200$. We {then} classify these $N_{p}=1000$ trajectories as predictable, and conversely, the $N_{p}=1000$ MFE trajectories that have the smallest averaged $\kappa$ between $t_{p} < t < t_{d}$ as unpredictable.

\begin{figure}
\centering
\includegraphics[width=\textwidth, trim=0mm 0mm 0mm 0mm, clip=true]{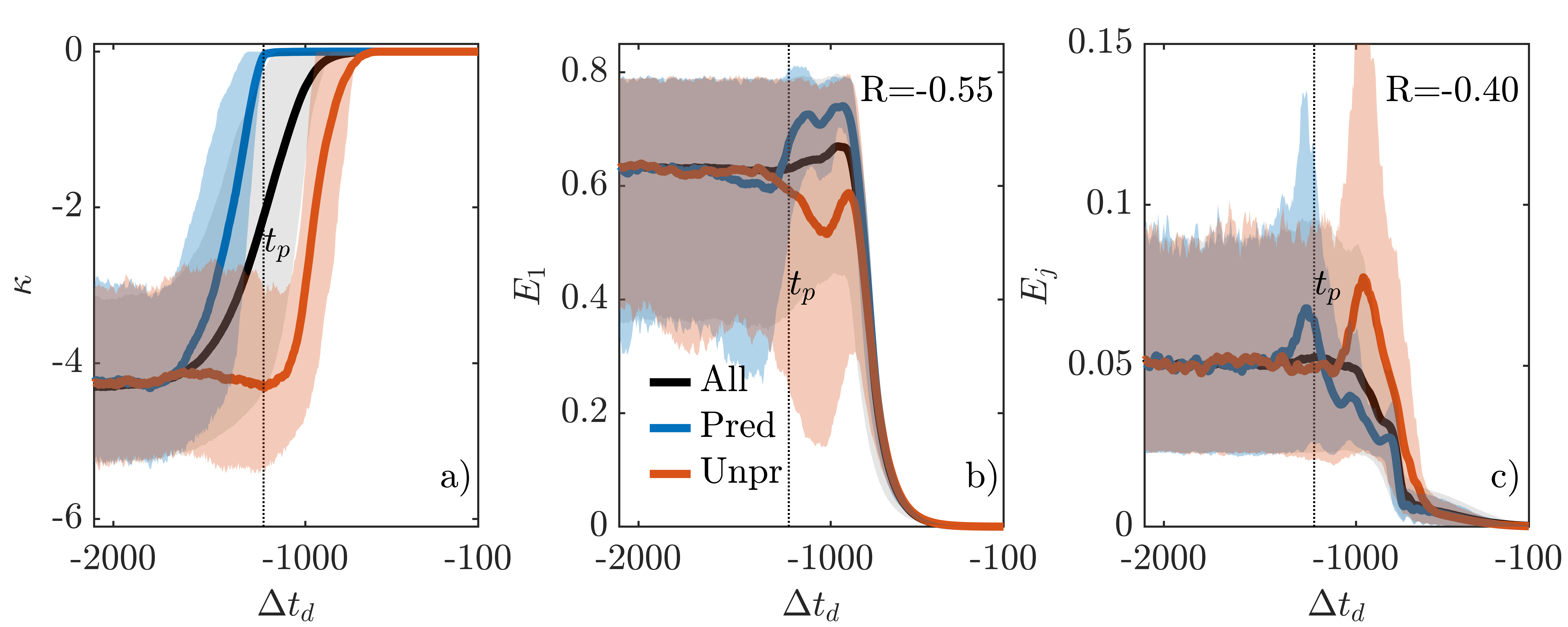}
\caption{Classification of MFE trajectories according to their predictability of decay. We consider three trajectory groups: all the trajectories $N_{b}=14000$ (in black) only the $N_{b}=1000$ most predictable trajectories (in blue), and only the $N_{b}=1000$ most unpredictable trajectories (in red) at $t_{p}<t<t_{d}$. Thick lines denote averaged quantities among the members of the group, and the shaded areas comprehend the first and last. a) predictability $\left(\kappa\right)$, eq.~\eqref{eq:kappa};  b) energy of the mean profile $E_{1}$, eq.~\eqref{eq:CH2:E1}; c) energy of the fluctuations $E_{j}$, eq.~\eqref{eq:CH2:Ej}. The vertical line denotes the threshold in time $t_{p}$ we use to differentiate between predictable and unpredictable trajectories. R stands for the temporal correlation between the plotted variable and $\kappa$, averaged among all the MFE trajectories. }
\label{fig:CH4:DMFET1}
\end{figure}
In figure~\ref{fig:CH4:DMFET1}a, we show predictability statistics of all the MFE trajectories (black), conditioned to either predictable (blue) or unpredictable trajectories (red). We fit each mean predictability with equation~\eqref{eq:CH3:fitKLD}, and analyze the resulting $t_{0}$ and $t_{c}$. As expected, $t_{0}$ is on average largest for the predictable trajectories, and smallest for the unpredictable trajectories. Between the two there is a difference of $ \Delta t_{0}= 452.12 $. We observe that $t_{c}$ is shorter, both for the predictable and unpredictable trajectories compared with all the data set. This means that, inside the predictable and unpredictable groups, the mechanisms behind decay are very similar among all the trajectories in the group, and that they take place in a relatively short time.

{We attempt to explain the two types of MFE decay events (predictable and unpredictable) from a fluid dynamics perspective. In figure~\ref{fig:CH4:DMFET1}b and \ref{fig:CH4:DMFET1}c, we show statistics of $E_{1}$ (energy of the mean profile, eq.~\eqref{eq:CH2:E1}) and $E_{j}$ (energy of the fluctuations, eq.~\eqref{eq:CH2:Ej}) of the three groups. The temporal correlation between $E_{1}$ (or $E_{j}$) with $K$, averaged among all the base trajectories, is negative. The predictable cases have a high amplitude $E_{j}$ phenomena at $t \approx t_{p}$. This event is subsequently followed by a flattening of the mean profile, $E_{1} \rightarrow 0.75$. In transitional shear flows a flat mean profile is linked with reduced turbulent fluctuations \citep{hof2010eliminating,barkley2015rise,kuhnen}. During the flattening of the profile, $E_{j}$ quickly decreases. At a certain time, $E_{1}$ starts to quickly decrease. Subsequently $E_{j}$ decreases in a non-monotonous way, but it never reaches a sufficiently high amplitude to re-trigger chaos. Before decaying $E_{j}$ shows some damped oscillations at $\Delta t_{d} \gtrsim -800$.}

The unpredictable cases, have a larger variance at all times $t_{p} < t < t_{d}$ than the predictable trajectories. This means that the identification of a clear mechanism of decay is more difficult for this group. Nevertheless, unpredictable trajectories have a peak $E_{j}$ event at $\Delta t_{d} \approx -1000$, that is followed by a rapid decrease of $E_{j}$. Although on average $E_{1}$ slightly increases after the $E_{j}$ peak, it quickly decreases together with $E_{j}$. Decay events for unpredictable trajectories have a sudden collapse of all the variables at almost the same time. They all quickly decay without showing damped oscillations as $t \rightarrow t_{d}$.

\subsection{Regions of the reduced order model phase space according to predictability}\label{sec:MFEregions}
\begin{figure}
\centering
\includegraphics[width=\textwidth, trim=0mm 0mm 0mm 0mm, clip=true]{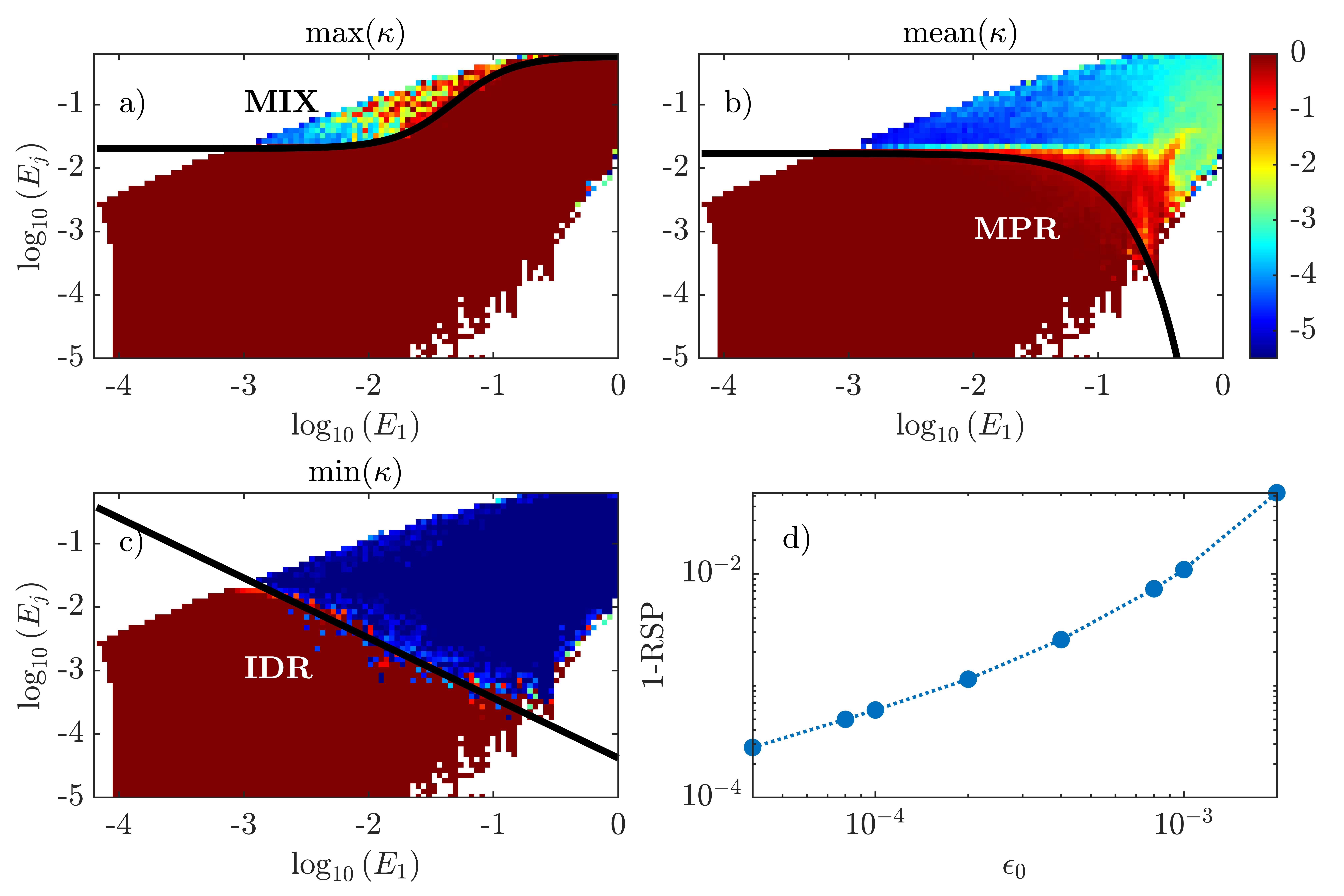}
\caption{Regions of the projected phase space of the MFE model depending on their predictability with respect to decay events. The sampled states of the base trajectories are projected in the $\log_{1} \left(E_{1}\right)$ and $\log_{10} \left( E_{j}\right)$ reduced phase space, that is divided in $100 \times 100$ bins of equal sizes. The color of each bin depends on the predictability of the states inside of it: red means high predictability (high $\kappa$), blue low (low $\kappa$). a) the color of each bin corresponds to the maximum predictability among all the states inside of it; b) to the mean and c) to the minimum. The solid black lines separate regions of phase space depending on their predictability. d) quality of the classification of cases in the IDR as one minus the ratio of successful predictions and its dependence on the noise magnitude $\epsilon_{0}$.}
\label{fig:CH5:DMFE}
\end{figure}

We project our MFE database in the $\log_{10} \left(E_{1}\right)$ and $\log_{10}\left( E_{j} \right)$ plane. We then divide the projected phase space in bins of equal sizes. In figure~\ref{fig:CH5:DMFE}a, we color the bins with the maximum $\kappa$ among all the instantaneous states in the bin. The bins colored in red, have at least one member inside that is highly predictable to decay. It is easy to recognize a big red region of phase space. This region occupies almost all the projected phase space where we have data points, showing that almost in any region of phase space there is at least one trajectory that quickly decays. In terms of dynamical systems this demonstrates that the edge is dense in the chaotic saddle \citep{budanur2019geometry}. However, there is a region in the plot, at high values of $\log_{10}\left(E_{j}\right)$, not colored in red. The bins in this region of the plot, only have instantaneous states that have a small predictability. This means that in this region decay is unpredictable. Cases that fall in this region are expected to (on average) remain chaotic for a relatively long time span. We call this region the Mixing Region (MIX). We find that the formula:
\begin{equation}
y_\text{MIX} \approx -0.96+0.73 \cdot \tanh \left\{ 2.1 \cdot \left[ \log_{10}\left(E_{1}\right) - 1.3 \right] \right\} \text{,}
\label{eq:CH5:yMIX}
\end{equation}
separates reasonably well this region from the rest of the projected phase space. At a given $\log_{10}\left(E_{1}\right)$, states with $\log_{10} \left( E_{j} \right)>y_\text{MIX}$, fall in the MIX Region.

In figure~\ref{fig:CH5:DMFE}b, we color the bins with the mean $\kappa$ among all the instantaneous states in the bin. There is a big region of the phase space colored in red, where the average predictability of decay is very high. We call it the Mean Predictable Region (MPR) and separate it from the rest of the projected phase space with the curve
\begin{equation}
y_\text{MPR} \approx -22.1+20.4 \cdot \tanh \left\{ -1.5 \cdot \left[ \log_{10}\left(E_{1}\right) - 0.5 \right] \right\} \text{.}
\label{eq:CH5:yMPR}
\end{equation}
At a given $\log_{10}\left(E_{1}\right)$, states with $\log_{10} \left( E_{j} \right)<y_\text{MPR}$, fall in the MPR. In figure~\ref{fig:CH5:DMFE}c, we color the bins with the minimum $\kappa$ among all the instantaneous states classified in that bin. The region in blue represents bins where at least one case classified in that bin is unpredictable to decay. The region in red, {referred to as the Inevitable Decay Region (IDR)}, corresponds to bins where all the cases classified inside are highly predictable to decay. Cases found in this region will almost certainly decay in a short time. We found that the best classifier, between the blue and red regions, is the line:
\begin{equation}
y_\text{IDR} \approx -0.95 \log_{10}\left(E_{1}\right) -4.4 \text{,}
\label{eq:CH5:yIDR}
\end{equation}
computed using a linear support vector machine \citep{cortes1995support}. At a given $\log_{10}\left(E_{1}\right)$ states with $\log_{10} \left( E_{j} \right)<y_\text{IDR}$, are almost guaranteed to decay in a short time. We recompute the regions of phase space according to the minimum $K$ for several uncertainty magnitudes $\epsilon_{0}\in \left[4\cdot 10^{-5},2 \cdot 10^{-3}\right]$, and find that the separating line is unaffected by $\epsilon_{0}$. We test the quality of the classification by computing a rate of successful predictions (RSP). We define the RSP as the quotient between the number of cases classified in the IDR that have $K \geq 0.99$ with respect to the total number of cases classified in the IDR. In case $\text{RSP}=1$, it means that all the cases in the IDR are fully predictable to decay. In figure~\ref{fig:CH5:DMFE}d we show the behavior of RSP with respect to $\epsilon_{0}$. We observe that, as $\epsilon_{0}$ decreases, the classification significantly improves. At $\epsilon_{0}=10^{-3}$, $1\%$ of the cases are wrongly classified in the IDR, while, at $\epsilon_{0}=10^{-4}$, only $0.1\%$ are. {We note that} RSP will never reach $\text{RSP} \rightarrow 1$. As we show later, there are extremely rare trajectories that can enter the IDR and subsequently remain chaotic for longer times. 

\subsection{A predictor of decay for the MFE model}
We use the different predictability regions discussed above to develop a simple predictive model of decay events in the MFE model. The predictive model only needs two inputs: the region in the $\left( E_{1},E_{j} \right)$ plane in which the trajectory is currently found in, and how long it has resided in that region. The model returns an indicator $\text{Ind}$ that is maximum $\left( \text{Ind}=1 \right)$ if decay is fully predictable, and smaller $0 \leq \text{Ind} < 1$ otherwise. See appendix~\ref{sec:Predictor} for more details. Our model correctly predicts decay (or not decay) more than $99 \%$ of the times. 

\begin{figure}
\centering
\includegraphics[width=\textwidth, trim=0mm 0mm 0mm 0mm, clip=true]{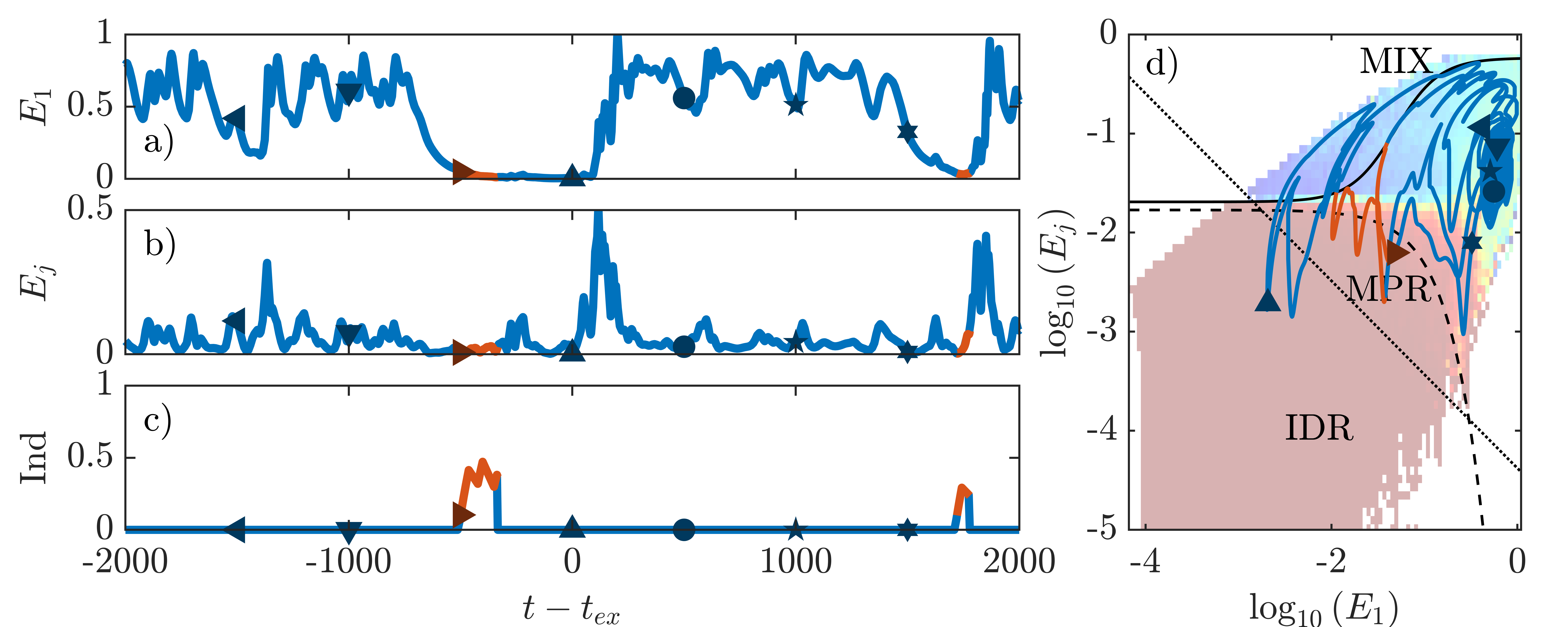}
\caption{{Rare MFE trajectory reaching $E=E_{1}+E_{j}\leq 0.004$ at $t=t_{ex}$ and with $t_{d} > t_{ex}+2000$. a) $E_{1}$ b) $E_{j}$ and c) our decay indicator Ind. Ind$=0$, no imminent decay is predicted; Ind$\gtrsim 0.8$ decay is imminent,} d) projection of the trajectory in the $\log_{10} \left(E_{1} \right)$, $\log_{10} \left(E_{j} \right)$ plane. The black lines separate the regions of predictability discussed in section \ref{sec:MFEregions} {(MIX with solid line, MPR with dashed, IDR with dotted)}. In all the plots, red means predictable, blue unpredictable, and the markers help to identify the moment in time.}
\label{fig:CH6:PredictorU1}
\end{figure}

We tested the predictor's performance with an extremely rare MFE trajectory {(diseminated in figure~\ref{fig:CH6:PredictorU1})}, which reaches an extremely low value of the kinetic energy:
\begin{equation}
E=E_{1} + E_{j} \leq 0.004 \text{,}
\end{equation}
at $t=t_{ex}$, {but remains chaotic for at least 2000 time units thereafter}. This event is similar to other extreme events observed in the MFE model, and described in appendix~\ref{app:XMFE}. We found this particular trajectory after trying billions of random initial conditions. 

We observe that this rare trajectory initially behaves chaotically in a small region of phase space, and then enters the MPR. It bounces back and forth, entering and exiting the MPR, oscillating about the $y_\text{MPR}$ line, as Ind increases. Even though we obtained the $y_\text{MPR}$ line from the perspective of {the mean} predictability, it is representative for the dynamics of this, and other trajectories (figure~\ref{fig:CH6:Predictor}). However, instead of decaying, it then enters the MIX region. As soon as a trajectory enters the MIX region, no predictions of decay are made for a time span of $\approx 2t_{c}$ time units. This is because, as discussed above, in this region states are highly unpredictable, and tend to decay after long times. Interestingly, after visiting the MIX region, the trajectory quickly crosses the MPR and IDR, and reaches the extremely low value of $E \leq 0.004$. Instead of decaying, however, it goes back to the MIX region and it then remains chaotic for a longer time, {in agreement with our predictor (Ind $=0$).}

This example shows that certain, albeit rare, trajectories can cross the IDR and still not subsequently decay. \cite{lellep2022interpreted} considered decay irreversible when $E \leq 0.005$. As we show here, trajectories can remain chaotic for long times after going below this threshold. We use this example to stress the importance of determining the threshold of decay for a level of allowed uncertainty, and the importance of characterizing regions of phase space according to predictability.

\section{Predictability of puff decay events}\label{sec:Puffresults}

\subsection{Classification of puff trajectories according to predictability}
\begin{figure}
\centering
\includegraphics[width=\textwidth, trim=0mm 0mm 0mm 0mm, clip=true]{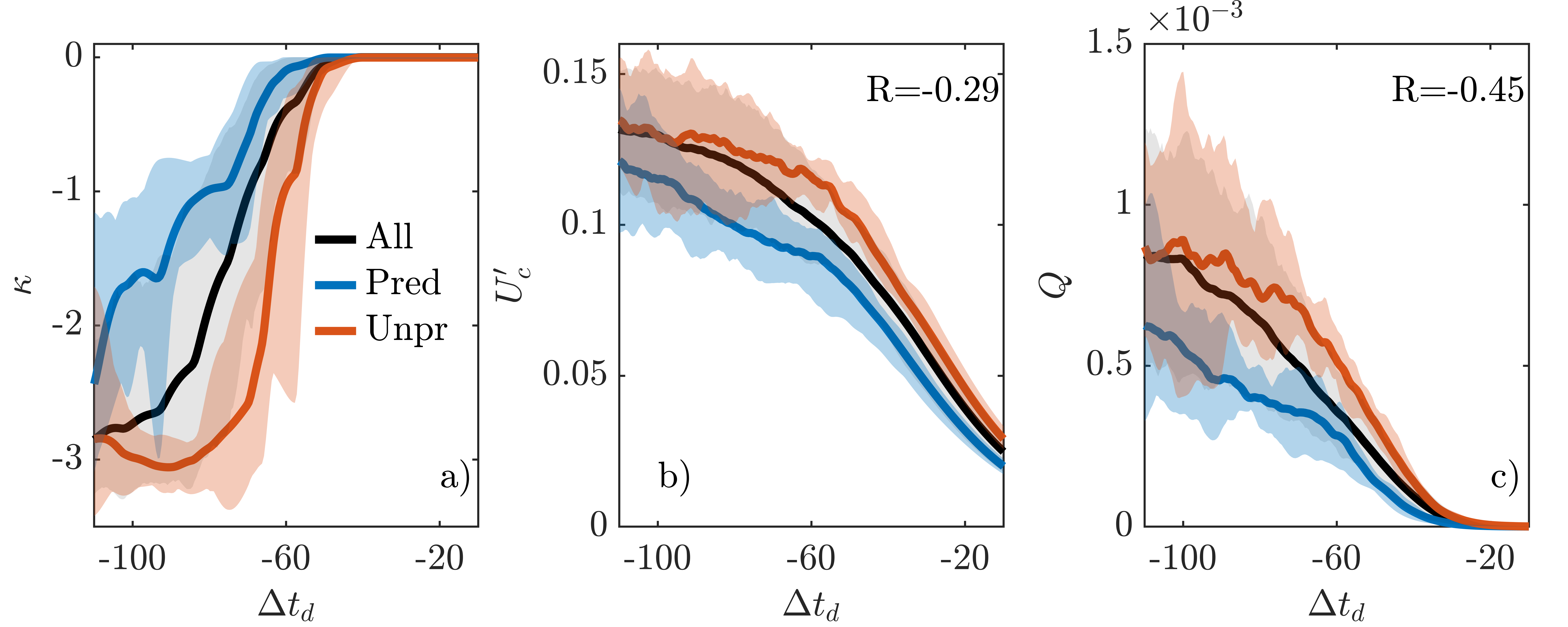}
\caption{Classification of puff trajectories according to their predictability of decay. We consider three trajectory groups: all the trajectories $N_{b}=100$ (in black) only the $N_{b}=10$ most predictable trajectories (in blue), and only the $N_{b}=10$ most unpredictable trajectories (in red). In all the plots the thick lines denote averaged quantities among the members of the group, and the shaded areas comprehend the first and last decile of each group. a) predictability ($\kappa$, eq.~\eqref{eq:kappa}); b) the volume averaged deviation from the center line velocity, eq.~\eqref{eq:CH2:Uc} and c) the volume averaged cross section kinetic energy, eq.~\eqref{eq:CH2:Q}. R stands for the time correlation between $\kappa$ and $Q$ or $U_{c}'$, averaged among all trajectories.}
\label{fig:CH4:PuffTrj1}
\end{figure}
We consider as predictable (unpredictable) the $N_{p}=10$ trajectories that have the largest (smallest) time averaged $\kappa$, eq.~\eqref{eq:kappa}, between $t_{p} < t < t_{d}$, for $t_{p}=t_{d}-120$. In figure~\ref{fig:CH4:PuffTrj1}a, we show the behavior of $\kappa$ for the three groups. We fit the averaged $\kappa$ of each group to equation~\eqref{eq:CH3:fitKLD}. As expected $t_{0}$ is larger for the predictable group, and smaller for the unpredictable group. We report here a difference of $\Delta t_{0}=23.52$ between the two. 

In contrast to the results of the MFE model, $t_{c}$ is much larger for the predictable group, $t_{c} \approx 28.96$, than for the unpredictable group $t_{c} \approx 8.47$. {A smaller $t_{c}$ implies that the group of trajectories have similar mechanisms of turbulence decay and that these mechanisms happen in a very short time, in an abrupt way. A large $t_{c}$ corresponds to trajectories that undergo a more gradual decay process, and whose mechanisms of decay are more different between each other.}

In figure~\ref{fig:CH4:PuffTrj1}b we show the behavior of the defect of center line velocity $U_{c}'$, eq.~\eqref{eq:CH2:Uc}, with respect to time. As the decay event is approached, the mean profile becomes more similar to the laminar one $\left(U_{c}' \rightarrow 0 \right)$. We observe that the predictable trajectories have, on average, a smaller value of $U_{c}'$ (corresponding to a more laminar like profile) than the unpredictable ones at all times. In figure~\ref{fig:CH4:PuffTrj1}c, we show the cross-section kinetic energy $Q$, eq.~\eqref{eq:CH2:Q}, with respect to time. {As the decay event approaches, $Q$ decreases, and it is always smaller on average in the case of the predictable trajectories. Both $U_{c}'$ and $Q$ correlate negatively with the predictability, as seen with the parameter R in the plots, and $Q$ has a better averaged temporal correlation.}

We note that, when full predictability is reached at $\Delta t_{d} \approx -40$, there is still a big uncertainty on the values of $Q$ and $U_{c}'$ a trajectory can have.  

\subsection{Regions of pipe flow phase space according to predictability}\label{sec:Puffregions}
\begin{figure}
\centering
\includegraphics[width=\textwidth, trim=0mm 0mm 0mm 0mm, clip=true]{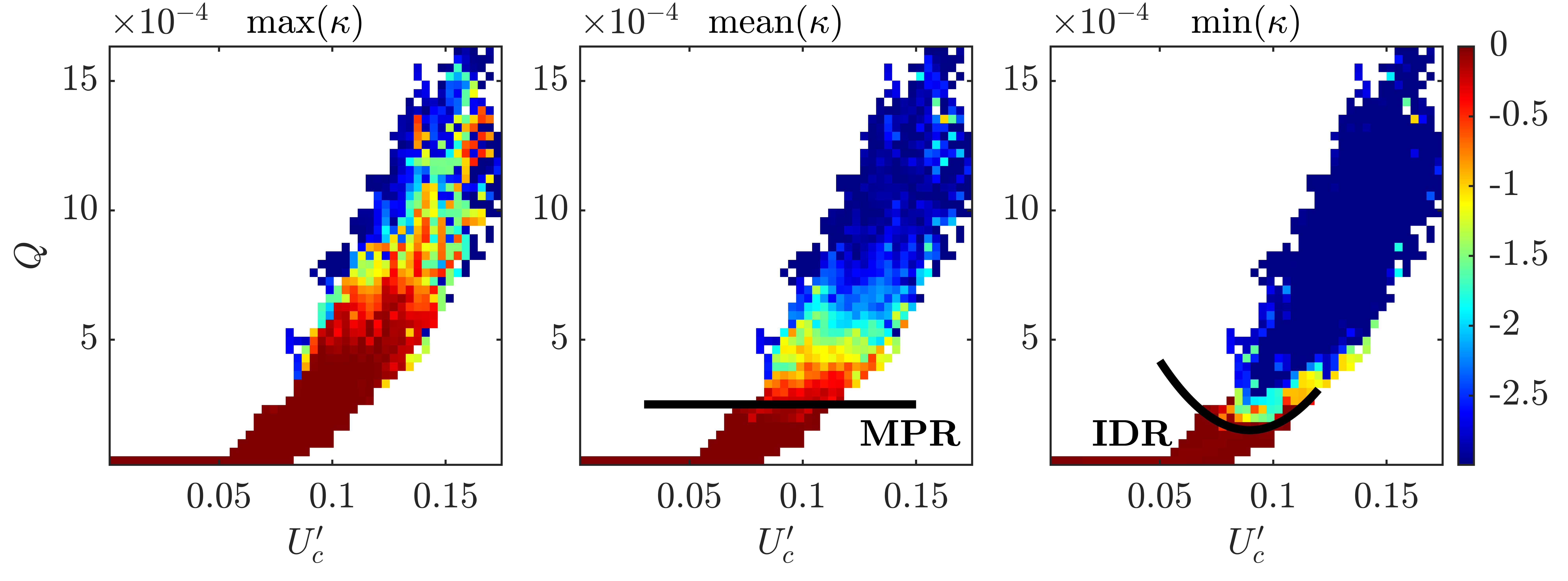}
\caption{Regions of the projected phase space of puffs, depending on their predictability with respect to decay events. The sampled states of the base trajectories are projected in the $U_{c}'$ and $Q$ reduced phase space, that is divided in $50 \times 50$ bins of equal sizes. The color of each bin depends on the predictability of the states inside of it: red means high predictability (high $\kappa$, eq.~\ref{eq:kappa}), blue low . a) the color of each bin corresponds to the maximum predictability of all the states inside of it, in b) to the mean and in c) to the minimum. The solid black line in b) indicates the mean predictable region (MPR) and in c) the inevitable decay region (IDR) of puffs, eq.~\eqref{eq:zIDR}. }
\label{fig:CH5:PuffR1}
\end{figure}

We project our puff database onto the $Q$ and $U_{c}'$ plane. We then divide the projected phase space in bins of equal sizes. In figure~\ref{fig:CH5:PuffR1}a we show the case of coloring each bin with the maximum $\kappa$ {among the states classified in that bin}. We observe a big red region that corresponds to bins where at least one case inside is highly predictable to decay. This region covers a huge portion of the sampled phase space, particularly for any $Q \lesssim 5\cdot 10^{-4}$. It is only for large excursions of $Q$ {that} puffs become highly unpredictable to decay. In figure~\ref{fig:CH5:PuffR1}b, we color each bin with the mean $\kappa$. We observe a red region, that is almost completely defined by the line $Q \lesssim 2.5\cdot 10^{-4}$. {Above this line} predictability gradually decreases for larger values of $Q$, {and we refer to the phase space below as the Mean Predictable Region of puffs.}

In figure~\ref{fig:CH5:PuffR1}c, we color each bin with the minimum $\kappa$. There is a clear red region at the bottom of the plot that represents the inevitable decay region of puffs. We find that one can separate this region with a parabola:
\begin{equation}
Q_\text{IDR} \approx 0.17~U_{c}'^{2} - 0.031~U_{c}' + 0.0015 \text{.}
\label{eq:zIDR}
\end{equation} 

There is an intermediate value of $U_{c}' \approx 0.09$, where $Q_\text{IDR}$ is minimum $Q_\text{IDR} \approx 1.51 \cdot 10^{-4}$. {Accordingly,} there are puffs that, despite reaching such a low $Q$ value, remain subsequently turbulent for longer times. 

At high $U_{c}' \gtrsim 0.1$ unless $Q$ is high enough, the flow will decay. This region corresponds to flows {with a relatively flat mean profile}. As we also saw for the MFE model, such flows need a high enough $Q$ to remain chaotic. 

Interestingly, for $U_{c}' \lesssim 0.07$, $Q$ must be high enough so the flow does not decay. At these low $U_{c}'$ the mean profile is very similar to the laminar one. One would expect a lower $Q$ threshold in this region, as fluctuations are more likely to proliferate using the laminar-like mean profile. This, however, is not the case. We speculate that, at these $U_{c}'$, fluctuations, no matter how intense they are, tend to be at wall-normal locations that do not promote turbulence survival, resulting in a high $Q_\text{IDR}$ threshold.

%We observe that states with $Q \approx Q_\text{IDR}$ are related with axial vorticity magnitudes far larger than the threshold identified by \cite{Alexakis}. We observe that all of our base trajectories reach their threshold far after they become fully predictable to decay.

%We theorize that the mean time of predictability loss does not depend on the event of interest, and represents an intrinsic measurement of the dynamical system. In fact, as we show in appendix~\ref{app:XMFE}, we obtain a similar $t_{c}$ after studying the predictability of a different, but also rare, (extreme) event in the MFE. Interestingly, in a similar study, \cite{lellep2022interpreted} reported a degradation of their predictor of MFE decay events, after a time span of a similar length to $t_{c}$. 

%%%%%%%%%%%%%%%%%%%%%%%%%%%%%%%%%%%%%%%%%%%%%%%%%%%%%%%%%%%%%%%%%
%                        CONCLUSION                             %
%%%%%%%%%%%%%%%%%%%%%%%%%%%%%%%%%%%%%%%%%%%%%%%%%%%%%%%%%%%%%%%%%
\section{Conclusions} \label{sec:Conclusions}
We propose a novel, probabilistic perspective to study the threshold of decay events in transitional shear flows. Instead of directly looking for the causes behind decay events, we first characterize their predictability using massive ensembles of simulations, and then identify the flow configurations that are highly predictable to decay. Our results show that all puff and MFE trajectories saturate to a maximum predictability at a given moment in time before decay {($\Delta t_{d} \approx -800 \text{ and }-40$ for the MFE model and puffs respectively)}. This result answers part of the question we initially pose, as to what is the first moment in time at which decay becomes inevitable. 

By analyzing the change of predictability with respect to time, we find a characteristic time of predictability loss $t_{c}$. Its inverse, $1/t_{c}$ is the rate at which predictability degrades, {on average}, as one goes back in time with respect to decay events. As we show {in appendix~\ref{app:robust}}, this measurement is almost unaffected by the level of uncertainty $\epsilon_{0}$ and, {as we show in appendix~\ref{app:XMFE}, it} does not depend on the (rare) event of interest: it is {an intrinsic} characteristic of the system. This has profound implications on the development of predictors of decay, and more generally of rare events, as the ability of any predictor will inevitably degrade exponentially, in time spans of the order of $t_{c}$. 

We identify the mechanisms by which two types of MFE trajectories decay. Trajectories that become fully predictable long before decay, have a sudden flattening of the mean profile that results in an abrupt collapse of the chaotic behavior and a decay behavior similar to a damped oscillator. Trajectories that {remain} unpredictable for {longer} times {before decay}, have a {sudden} high amplitude {fluctuation} event, after which turbulence quickly collapses and becomes laminar without oscillations. By projecting our data in a two-dimensional phase space, we identify different regions depending on their predictability. We report a region where all the states are highly unpredictable, a region where most cases are predictable and a region where (almost) all cases are fully predictable to decay. The last region answers part of our initial question, {as to}: what flow configurations are fully predictable to (and therefore are a threshold of) decay. We use these regions of predictability to develop a simple predictive model of MFE decay, that returns a high ratio of successful predictions {RSP $>0.99$}.  

{We repeat the above analysis for the case of puffs in pipe flow, and consider two key variables defined by \cite{barkley2011simplifying}, and thereafter widely used to study transitional pipe flow: the kinetic energy $Q$ of cross flow fluctuations, and the deviation from the laminar center line velocity, $U_{c}'$. We identify a region of phase space that is fully predictable to decay, and report, for the first time, a threshold of puff decay in terms of $Q$ and $U_{c}'$.} We observe that the threshold of decay is mostly set by a low value of $Q$, in line with previous studies \citep{Alexakis}, but depends slightly on $U_{c}'$. We corroborate that a pipe flow with a flat mean profile needs a high $Q$ to survive \citep{kuhnen}. Interestingly we also observe the opposite, flows with a mean profile similar to the laminar one, need also a high $Q$ to survive. We speculate that, for these flows, the fluctuations may be found at locations far from the wall that do not promote turbulence survival. 

{We argue that the study of predictability can ultimately identify the causes of relaminarisation. Specifically a flow configuration is causal to relaminarisation to the same extent the latter is predictable from the former. It is our objective to, in future analyses, look for flow variables that better correlate with predictability, and therefore, with the causes of decay. A promising prospect is the combination of our probabilistic approach, with the study of invariant solutions in shear flows \citep{kawahara2012significance}. In future analyses we aim to use invariant solutions of pipe flow at the edge of chaos (e.g. \cite{duguet2008transition}), as initial conditions for our ensembles. We believe that, by doing this, we can identify regions of phase space that are more/less attracting to these invariant solutions and exploit their characteristics to better understand the dynamics of decay events.}
\\
\\
\textbf{Acknowledgments.}  Part of this work was done during the Fifth Madrid Summer Workshop, funded by the European Research Council (ERC) under the Caust grant ERC-AdG-101018287. The members and organizers of the workshop are here greatly acknowledged. The authors would also like to thank Dr. Miguel P. Encinar for fruitful discussions.
\\
\\
\textbf{Declaration of interest.} The authors report no conflict of interest.
\\
\\
\textbf{Data availability statement.} All the codes and data are available in Pangaea under the following \href{https://doi.pangaea.de/10.1594/PANGAEA.977819}{link}. The GPU code to simulate pipe flow can be obtained in the following Github: \href{https://github.com/Mordered/nsPipe-GPU.git}{https://github.com/Mordered/nsPipe-GPU.git}.

\appendix

\section{Robustness of the method to compute predictability}\label{app:robust}
\begin{figure}
\centering
\includegraphics[width=\textwidth, trim=0mm 0mm 0mm 0mm, clip=true]{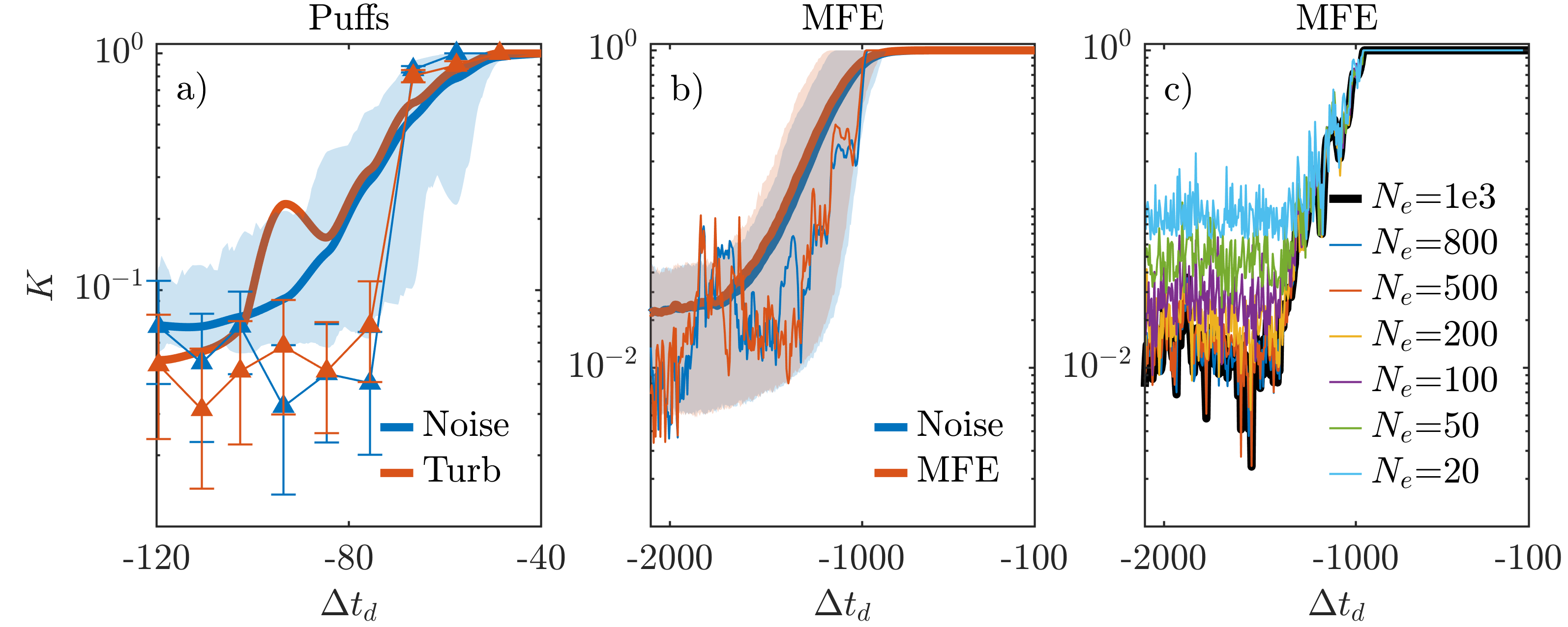}
\caption{Robustness of the predictability measurement. a) predictability of puff decay events depending on the type of initial condition used in the ensembles of simulations. Blue corresponds to random Gaussian noise with $\epsilon_{0} \approx 10^{-2}$, red to a scaled fully turbulent field. The thick lines correspond to the mean predictability: averaged over $N_{b}=100$ base trajectories, for the case of random noise, and of only $N_{b}=4$ base trajectories for the case of scaled turbulent fields. The thin lines correspond to the same base trajectory: in red with a predictability characterized with the scaled turbulent field, in blue with random noise. b) predictability of decay events in the MFE model, for ensembles initialized with Gaussian noise (blue), and ensembles initialized with a scaled MFE chaotic state (red). c) predictability with respect to time of a single MFE decay trajectory, characterized with ensembles of different sizes $N_{e}$.}
\label{fig:CH3:IniCnd}
\end{figure}

\subsection{Effect of the shape of the uncertainties on predictability}
Here we explore the effect of using a different type of perturbation than Gaussian noise to initialize the ensembles of simulations. In the case of puff decay, we performed a simulation with the same discretization described in section \ref{sec:Puffsmethods}, but at $Re=3000$. We trigger turbulence in the simulation and wait until the whole domain is fully turbulent. We then save perturbed velocity fields $\pmb{u'}=\pmb{u}-\left(0,0,u_{HP}\left( r\right)\right)$ at $N_{e}$ time steps, after every $1.25$ advective time unit. We initialize each member of the ensemble with one of the scaled turbulent fields as:
\begin{equation}
\pmb{u} \left(t=0\right)= \pmb{u}_{t_{k}}^{i} + \epsilon_{T} \frac{\pmb{u'}}{\left| \pmb{u'} \right|} \text{.}
\end{equation}
Note that this type of perturbation satisfies continuity of the flow. We fix $\epsilon_{T}=2.5\cdot 10^{-2}$ to make the energy of the perturbation equivalent to the energy of the Gaussian perturbation at $\epsilon_{0} \approx 10^{-2}$. We re-compute ensembles of simulations using $N_{t}=10$ instantaneous states of only $N_{b}=4$ of our base trajectories, see table~\ref{tab:CH3:param}. In figure~\ref{fig:CH3:IniCnd}a we show predictability with respect to time depending on the type of perturbation used for the ensembles. We observe that the type of perturbation has little effect on the predictability of individual trajectories. Even after averaging over only $N_{b}=4$ trajectories, the mean trends of predictability remain unaffected.

For the case of the MFE decay events, we save $N_{e}$ instantaneous chaotic states of the MFE model at $t<t_{d}-2000$. We then re-scale the states to a magnitude of $\epsilon_{0}=10^{-4}$ and use them to initialize ensembles of simulations for $N_{b}=2000$ trajectories. In figure~\ref{fig:CH3:IniCnd}b we show predictability with respect to time depending on the type of perturbation used for the ensembles. We observe that the shape of the initial condition does not have an important impact on the mean trends of predictability, on the statistics or even on the predictability of individual trajectories.

\subsection{Effect of using less members per ensemble}
We also consider here the effect of using smaller members per ensemble $N_{e}$ on the determination of the predictability of MFE decay events. We use the case $N_{e}=1000$ as the base case scenario, and then reduce $N_{e}$ to study the changes on $K$, see fig.~\ref{fig:CH3:IniCnd}c. We observe that, by decreasing $N_{e}$, we obtain more noisy predictability estimations, specially for small values of $K$. This is expected as we have a larger uncertainty in our statistics. Nevertheless we report that the mean trends of predictability are well captured for all the $N_{e}$ considered here. Even the more limiting case of $N_{e}=20$ is able to capture the high amplitude oscillations of predictability observed at $N_{e}=1000$. We repeat this analysis for additional $N_{b}$ trajectories, compute the ensemble-averaged predictability and fit it to equation~\ref{eq:CH3:fitKLD}. We do this for all $N_{e}$. Although we do not show it here, we observe that between $N_{e}=20$ and $N_{e}=1000$, the fitted $t_{c}$ only changes in a $15 \%$, while between $N_{e}=200$ and $N_{e}=1000$ less than a $2 \%$. Thus, ensembles with small members result in almost identical predictability characteristics than at $N_{e}=1000$.

\subsection{The effect of the magnitude of uncertainties $\epsilon_{0}$ on predictability}\label{sec:CH3:ep0}
\begin{figure}
\centering
\includegraphics[width=\textwidth, trim=0mm 0mm 0mm 0mm, clip=true]{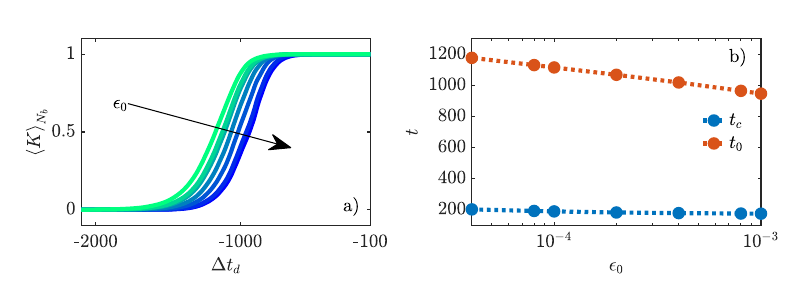}
\caption{Effect of the magnitude of uncertainties $\epsilon_{0}$ on predictability of MFE decay. a) mean $K$ with respect to time among $N_{b}=14000$ MFE base trajectories, computed with different magnitudes of the Gaussian noise. Lighter color means smaller $\epsilon_{0}$. b) results of fitting mean $K$ of different $\epsilon_{0}$ with equation~\eqref{eq:CH3:fitKLD}.}
\label{fig:CH3:ep0KLD}
\end{figure}
We recompute the predictability of the $N_{b}=14000$ decaying MFE trajectories, sampled at $N_{t}=400$ time steps and perturbed with a Gaussian noise, see table~\ref{tab:CH3:param}, but with different noise magnitudes: $\epsilon_{0}\in \left[4\cdot 10^{-5}, 2 \cdot 10^{-3}\right]$. In figure~\ref{fig:CH3:ep0KLD} we show the mean predictability for each $\epsilon_{0}$ and observe that, despite a certain horizontal shift, they are almost identical.

We fit the mean predictability to equation~\eqref{eq:CH3:fitKLD}. We report that, by decreasing $\epsilon_{0}$, the bias $t_{0}$ increases $t_{0} \propto - \log\left(\epsilon_{0}\right)$. In our analysis, $\epsilon_{0}$ models the level of uncertainty one has (e.g. from measurement device, measurement method, numerical method, ...). As it decreases, the bias increases as expected: with less uncertainties, one can predict decay earlier. Note that at $\epsilon_{0}=0$, $t_{0}\rightarrow \infty$. 

Although we observe a slight increase of $t_{c}$ as $\epsilon_{0}$ decreases, the characteristic time of predictability loss is almost unaffected by the size of $\epsilon_{0}$. Here we report a difference  of $\approx 13 \%$ between the two limiting $\epsilon_{0}$ cases $\epsilon_{0}=2 \cdot 10^{-3}$ and $\epsilon_{0}=4\cdot 10^{-5}$. This further suggests that $t_{c}$ is an intrinsic measurement of the dynamical system: it does not sensibly depend on the coarse graining that one chooses.

\section{Rare events in the reduced order model}\label{app:XMFE}
\begin{figure}
\centering
\includegraphics[width=\textwidth, trim=0mm 0mm 0mm 0mm, clip=true]{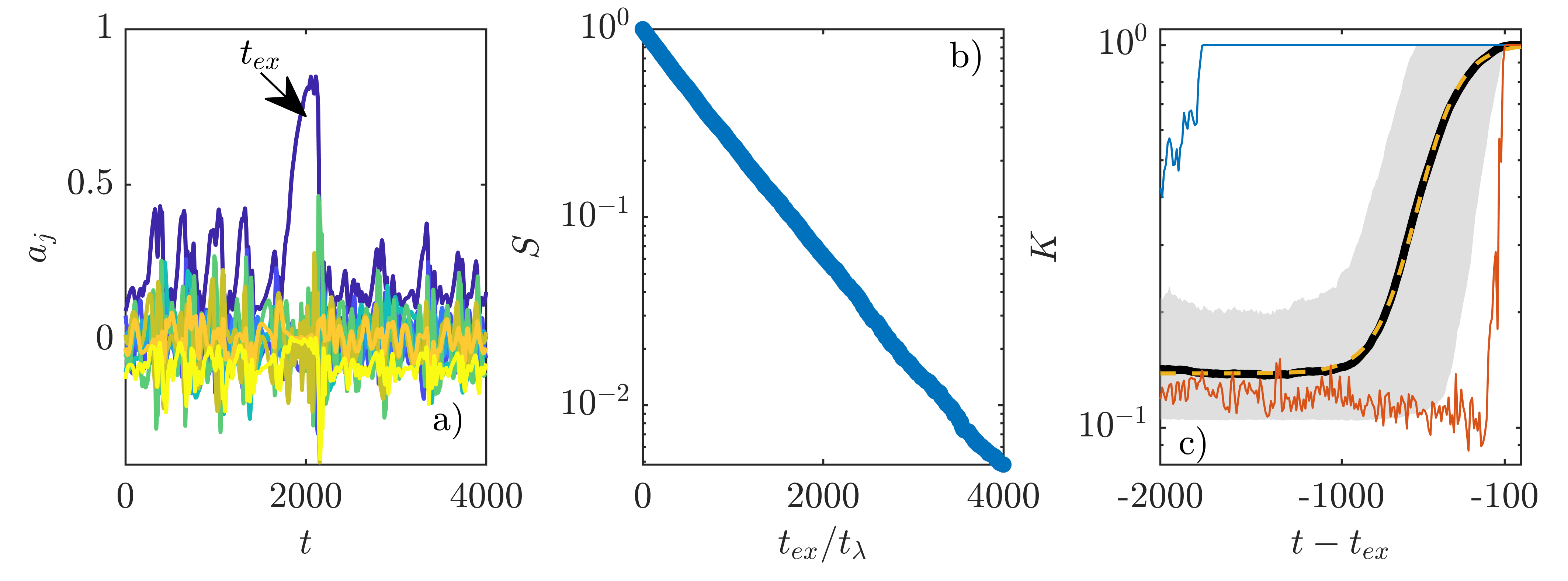}
\caption{Description and predictability of rare $a_{1}$ events in the MFE model. a) a MFE trajectory with an event of high $a_{1} \geq 0.8$ at time $t=t_{ex}$ and with $t_{d} > t_{ex}+2000$. b) statistics of this type of events in the MFE model. At $Re=400$ these events follow a Poisson distribution, with a mean waiting time $\tau_{ex}\approx 2.5 \cdot 10^{4}$. c) predictability of the MFE model to rare $a_{1}$ events. Thin lines correspond to the more (blue) and less (red) predictable trajectories, the shaded region comprehends the area between the first and last decile of the statistics and the thick line denotes the mean predictability. The dashed yellow line corresponds to the resultant fit of the mean $K$ to the formula in equation~\eqref{eq:CH3:fitKLD}.}
\label{fig:XMFE}
\end{figure}

At $Re=400$, at random times, the mode $a_{1}$ of the MFE model becomes very large $a_{1} \geq 0.8$, giving the impression that chaos may be about to decay. However, instead of subsequently decaying, the model then remains chaotic for relatively long times. See an example of a trajectory with such an event in figure\ref{fig:XMFE}a and also in figures~\ref{fig:fig1}c, \ref{fig:CH6:PredictorU1} and \ref{fig:CH6:Predictor}.

We call these events "extreme events" of the MFE and identify them at time $t=t_{ex}$ according to two conditions:
\begin{enumerate}
\item at $t=t_{ex}$, $a_{1}\geq 0.8$, 
\item and $t_{d} \geq t_{ex}+2000$.
\end{enumerate}
This means that, these events are highly intense on $a_{1}$, but do not quickly decay afterwards. The mean waiting time between these events follows an exponential distribution, as seen in figure~\ref{fig:XMFE}b. This means that, like the decay events, these events also follow a memoryless process.

As we did for the case of decay events, we compute the predictability of $N_{b}=10000$ base trajectories that have an extreme event at time $t_{ex}$. We sample the trajectories at $N_{t}=200$ time steps before the extreme event and launch ensembles of $N_{e}=1000$ simulations with $\epsilon_{0}=10^{-4}$ for each sampled state. We gather statistics of the times of extreme event recurrences for each ensemble, and compute $K$ as in equation~\eqref{eq:CH3:KLD}, but comparing the conditional distributions to the expected exponential distribution in figure~\ref{fig:XMFE}b.

We show the resultant predictability statistics in figure~\ref{fig:XMFE}c. As for the case of predictability of decay, predictability increases on average as one approaches the extreme event in time. We also observe differences in predictability of orders of magnitude between different trajectories. We fit the mean predictability to equation~\eqref{eq:CH3:fitKLD}. We report that the resultant $t_{c}=238.28$ is almost equal to the $t_{c} = 264.08$ computed for the predictability of decay events, see table~\ref{tab:CH3:timsca}. This suggests that the  predictability loss time scale is a general measurement of a chaotic system, and does not depend on the particular memoryless process of interest.

\section{Description of the predictor of MFE decays}\label{sec:Predictor}
\begin{figure}
\centering
\includegraphics[width=\textwidth, trim=0mm 0mm 0mm 0mm, clip=true]{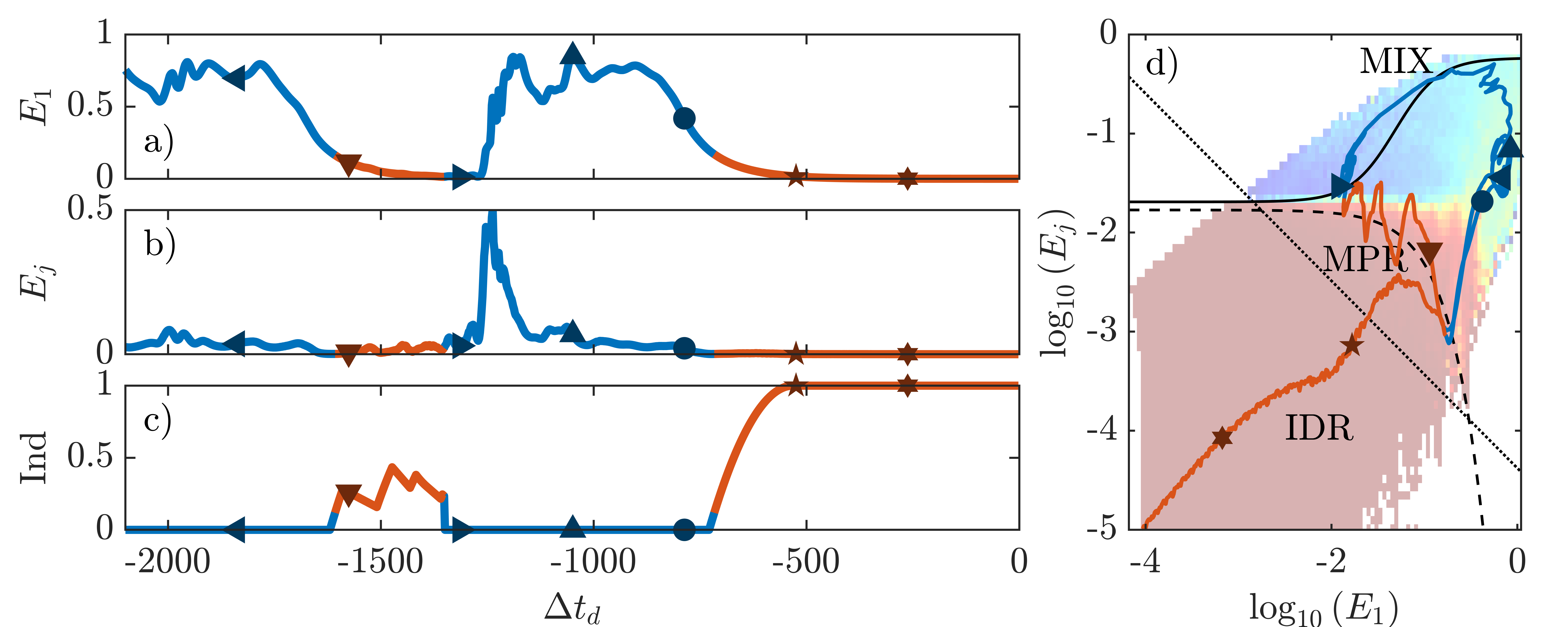}
\caption{{MFE trajectory, and our time dependent decay predictor. a) $E_{1}$ b) $E_{j}$ and c) our decay indicator Ind. Ind$=0$, no imminent decay is predicted; Ind$\gtrsim 0.8$ decay is imminent, d) projection of the trajectory in the $\log_{10} \left(E_{1} \right)$, $\log_{10} \left(E_{j} \right)$ plane.} The black lines separate the regions of predictability discussed in section \ref{sec:MFEregions} {(MIX with solid line, MPR with dashed, IDR with dotted)}. In all the plots, red means predictable, blue unpredictable, and the markers help to identify the moment in time.}
\label{fig:CH6:Predictor}
\end{figure}

The indicator of decay or not decay in our predictor model is computed as
\begin{equation}
\text{Ind}=\frac{\sigma}{t_\text{Ind}}  \int^t_{t-t_\text{Ind}}  \frac{1}{2} \left \{\text{sgn}\left[y_\text{MPR}-\log_{10}\left(E_{j}\right)\right]+1\right \} \left( \frac{2t}{t_\text{Ind}} \right) \mathrm{d}t \text{,}
\label{eq:CH6:PInd}
\end{equation}
where
\begin{equation}
\sigma =0 \text{ if } \int^t_{t-t_\text{MIX}}  \frac{1}{2} \left \{\text{sgn}\left[\log_{10}\left(E_{j}\right) - y_\text{MIX}\right]+1\right \} \mathrm{d}t > 0 \text{, and }  \sigma=1 \text{ otherwise,}
\end{equation}
being $\text{sgn}$ the signum function. The parameter $t_\text{Ind}$ is computed as:
\begin{equation}
t_\text{Ind}= \min \left(t_\text{MPR},t-t_{\sigma}\right) \text{,}
\end{equation}
being $t_{\sigma}$ the last time step where $\sigma<1$. 

Here, $y_\text{MIX}$ and $y_\text{MPR}$ are the delimiters of the mixing, eq.~\eqref{eq:CH5:yMIX}, and mean predictable regions, eq.~\eqref{eq:CH5:yMPR}; and $t_\text{MPR}$ and $t_\text{MIX}$ two parameters of the predictor. Here we tune $t_\text{MPR}=100$ and $t_\text{MIX}=400$ so they have the same order of magnitude as $t_{c}$.

When the MFE enters the MPR the indicator $\text{Ind}>0$, and increases in magnitude the longer the time the trajectory spends there. As soon as the time spent in the MPR is $t \geq t_\text{MPR}$, the indicator saturates to $1$. The predictor also checks if the trajectory enters the MIX region. In this region decay is highly unpredictable. As soon as the trajectory enters the MIX region, $\text{Ind}=0$ and no predictions are made for $t_\text{MIX}$ time units.

As long as $\text{Ind}>0$, the predictor also outputs a forecast decay time:
\begin{equation}
\Delta t_\text{decay} \left(t \right)=\frac{4 Re}{\pi^{2}}\log\left( \frac{1-a_{1}\left(t \right)}{1-0.995} \right) \text{,}
\end{equation}
That represents the forecasted remaining lifetime before decay. This decay time is derived from the equation of the variable $a_{1}$ \citep{moehlis2004low} by neglecting the nonlinear terms:
\begin{equation}
\frac{\mathrm{d}a_{1}}{\mathrm{d}t}= \frac{\pi^{2}}{4 Re}\left(1 -a_{1}\right) \text{,}
\label{eq:da1}
\end{equation}
integrating and assuming that decay is completed at $a_{1} \approx 0.995$ (our heuristic threshold). 

See in figure~\ref{fig:CH6:Predictor}c an example of the time dependent Ind for the MFE base trajectory shown in figure~\ref{fig:fig1}a. In figure~\ref{fig:CH6:Predictor}d we show the position of the trajectory in the projected phase space. We observe how the trajectory starts in the right top corner of the plot, where we observe most trajectories spend the majority of their time. The trajectory then starts to approach the MPR. As soon as the trajectory enters the MPR, Ind increases (represented by a change in color of the line).

The trajectory then enters the MIX region, and all predictions are discarded for a certain time span. Finally the trajectory enters again the MPR, and Ind increases monotonically to 1, as the trajectory finally decays.

\subsection{Ability of the predictor}
To asses the quality of our predictor, we define the function:
\begin{equation}
g\left( t \right)=\exp \left[- \left(\frac{\Delta t_\text{decay}\left(t\right) - t_{d} + t }{t_\text{tol}} \right)^{6}     \right] \text{,}
\end{equation}
where $t_\text{decay}$ is the forecasted remaining lifetime, $t_{d}$ the actual time to decay and $t_\text{tol}=800$. We set this tolerance to alleviate the strong assumption we make in equation~\eqref{eq:da1}, where we assume a viscous decay. The above function is equal to $\approx 1$ at $\left| \Delta t_\text{decay}\left(t\right) - t_{d} + t \right| \leq 400$ and smaller for any other time span:  $g \approx 0.5$ at $\left| \Delta t_\text{decay}\left(t\right) - t_{d} + t \right| \leq 750$. Note that, as shown in figure~\ref{fig:CH3:mKLD}b, all MFE trajectories reach maximum predictability at $t \approx 800$ time units before decay $t_{d}$.

We compute the error of the predictor with respect to time as the product:
\begin{equation}
\epsilon_{1}=\text{Ind} \cdot \left(1-g\right) \text{.}
\end{equation}
We compute the false positives of a base trajectory as:
\begin{equation}
\text{FP} = \frac{1}{\Delta t_\text{Ind}} \int_{0}^{\Delta t_\text{Ind}} \epsilon_{1} \mathrm{d} t \text{,}
\end{equation}
where $\Delta t_\text{Ind}$ are all the time steps where $\text{Ind}>0$. We compute the ratio of successful predictions as:
\begin{equation}
\text{RSP} = 1- \frac{1}{T} \int_{0}^{T} \epsilon_{1} \mathrm{d} t \text{,}
\end{equation}
to account for the amount of time the predictor is successfully predicting decay (or not decay). Here $T=2100$ is the total time span of each base trajectory. We compute the error and the RSP for all our $N_{b}=14000$ MFE base trajectories.

We report that the model returns a mean $\left \langle \text{FP} \right \rangle_{b}=7.16 \%$, out of which only $0.77 \%$ are full false positives. Full false positives are cases where $\epsilon_{1} \geq 0.95$. We report that, the model returns a mean  $\left \langle \text{RSP} \right \rangle_{b,t}=99.73 \%$, as it predicts correctly most of the time that the MFE is, or is not, about to decay.

\section{The MFE equations}\label{ap:MFEeq}
{We include in this appendix the MFE equations, as originally derived by \cite{moehlis2004low}. Let:
\begin{align*}
\beta = \frac{\pi}{2}\text{, }\alpha=\frac{2 \pi}{L_{x}} \text{, }\gamma=\frac{2 \pi}{L_{z}}\text{, }k_{\alpha\gamma}=\sqrt{\alpha^{2} + \gamma^{2}}\text{, }k_{\beta\gamma}=\sqrt{\beta^{2} + \gamma^{2}}\text{ and }k_{\alpha\beta\gamma}=\sqrt{\alpha^{2} + \beta^{2} + \gamma^{2}} \text{.}
\end{align*}
The equation for each mode reads:}
{\small
\begin{align*}
\frac{\mathrm{d}a_{1}}{\mathrm{d}t} &= \left( \frac{\beta^2}{Re} \right) (1 - a_1) 
+ \sqrt{1.5} \beta \gamma \left( \frac{a_2 a_3}{k_{\beta\gamma}} - \frac{a_6 a_8}{k_{\alpha\beta\gamma}} \right) 
\end{align*}
\begin{align*}
\frac{\mathrm{d}a_{2}}{\mathrm{d}t} &=- \frac{a_2}{Re} \left( \frac{4\beta^2}{3} + \gamma^2 \right)
+ a_4 a_6 \frac{\sqrt{50/27} \gamma^2}{k_{\alpha\gamma}}
- a_5 a_7 \frac{\gamma^2}{\sqrt{6} k_{\alpha\gamma}}\\
& - a_5 a_8 \frac{\alpha \beta \gamma}{\sqrt{6} k_{\alpha\gamma} k_{\alpha\beta\gamma}}
- a_1 a_3 \frac{\sqrt{1.5} \beta \gamma}{k_{\beta\gamma}}
- a_3 a_9 \frac{\sqrt{1.5} \beta \gamma}{k_{\beta\gamma}}
\end{align*}
\begin{align*}
\frac{\mathrm{d}a_{3}}{\mathrm{d}t} &= - a_3 \frac{\beta^2 + \gamma^2}{Re}
+ a_4 a_7 \frac{2 \alpha \beta \gamma}{\sqrt{6} k_{\alpha\gamma} k_{\beta\gamma}}
+ a_5 a_6 \frac{2 \alpha \beta \gamma}{\sqrt{6} k_{\alpha\gamma} k_{\beta\gamma}} + a_4 a_8 \frac{\beta^2 (3 \alpha^2 + \gamma^2) - 3 \gamma^2 (\alpha^2 + \gamma^2)}{\sqrt{6} k_{\alpha\gamma} k_{\beta\gamma} k_{\alpha\beta\gamma}}
\end{align*}
\begin{align*}
\frac{\mathrm{d}a_{4}}{\mathrm{d}t} &= - a_4 \frac{3\alpha^2 + 4\beta^2}{3 Re}
- a_1 a_5 \frac{\alpha}{\sqrt{6}}
- a_2 a_6 \frac{10 \alpha^2}{3 \sqrt{6} k_{\alpha\gamma}}
- a_3 a_7 \frac{\sqrt{1.5} \alpha \beta \gamma}{k_{\alpha\gamma} k_{\beta\gamma}}\\
& - a_3 a_8 \frac{\sqrt{1.5} \alpha^2 \beta^2}{k_{\alpha\gamma} k_{\beta\gamma} k_{\alpha\beta\gamma}}
- a_5 a_9 \frac{\alpha}{\sqrt{6}} 
\end{align*}
\begin{align*}
\frac{\mathrm{d}a_{5}}{\mathrm{d}t} &= - a_5 \frac{\alpha^2 + \beta^2}{Re}
+ a_1 a_4 \frac{\alpha}{\sqrt{6}}
+ a_2 a_7 \frac{\alpha^2}{\sqrt{6} k_{\alpha\gamma}}
- a_2 a_8 \frac{\alpha \beta \gamma}{\sqrt{6} k_{\alpha\gamma} k_{\alpha\beta\gamma}}
+ a_4 a_9 \frac{\alpha}{\sqrt{6}} + a_3 a_6 \frac{2 \alpha \beta \gamma}{\sqrt{6} k_{\alpha\gamma} k_{\beta\gamma}}
\end{align*}
\begin{align*}
\frac{\mathrm{d}a_{6}}{\mathrm{d}t} &= - a_6 \frac{3\alpha^2 + 4\beta^2 + 3\gamma^2}{3 Re}
+ a_1 a_7 \frac{\alpha}{\sqrt{6}}
+ a_1 a_8 \frac{\sqrt{1.5} \beta \gamma}{k_{\alpha\beta\gamma}}
+ a_2 a_4 \frac{10(\alpha^2 - \gamma^2)}{3 \sqrt{6} k_{\alpha\gamma}} \\
& - a_3 a_5 \frac{2 \alpha \beta \gamma}{\sqrt{1.5} k_{\alpha\gamma} k_{\beta\gamma}} + a_7 a_9 \frac{\alpha}{\sqrt{6}}
+ a_8 a_9 \frac{\sqrt{1.5} \beta \gamma}{k_{\alpha\beta\gamma}} 
\end{align*}
\begin{align*}
\frac{\mathrm{d}a_{7}}{\mathrm{d}t} &= - a_7 \frac{\alpha^2 + \beta^2 + \gamma^2}{Re}
- a_1 a_6 \frac{\alpha}{\sqrt{6}}
- a_6 a_9 \frac{\alpha}{\sqrt{6}}
+ a_2 a_5 \frac{\gamma^2 - \alpha^2}{\sqrt{6} k_{\alpha\gamma}}
+ a_3 a_4 \frac{\alpha \beta \gamma}{\sqrt{6} k_{\alpha\gamma} k_{\beta\gamma}}
\end{align*}
\begin{align*}
\frac{\mathrm{d}a_{8}}{\mathrm{d}t} &= - a_8 \frac{\alpha^2 + \beta^2 + \gamma^2}{Re}
+ a_2 a_5 \frac{2 \alpha \beta \gamma}{\sqrt{6} k_{\alpha\gamma} k_{\alpha\beta\gamma}}
+ a_3 a_4 \frac{\gamma^2 (3 \alpha^2 - \beta^2 + 3 \gamma^2)}{\sqrt{6} k_{\alpha\gamma} k_{\beta\gamma} k_{\alpha\beta\gamma}} 
\end{align*}
\begin{align*}
\frac{\mathrm{d}a_{9}}{\mathrm{d}t} &= - a_9 \frac{9 \beta^2}{Re}
+ a_2 a_3 \frac{\sqrt{1.5} \beta \gamma}{k_{\beta\gamma}}
- a_6 a_8 \frac{\sqrt{1.5} \beta \gamma}{k_{\alpha\beta\gamma}}
\end{align*}
}

%%%%%%%%%%%%%%%%%%%%%%%%%%%%%%%%%%%%%%%%%%%%%%%%%%%%%%%%%%%%%%%%%
%                       BIBLIOGRAPHY                            %
%%%%%%%%%%%%%%%%%%%%%%%%%%%%%%%%%%%%%%%%%%%%%%%%%%%%%%%%%%%%%%%%%
\bibliographystyle{jfm}
\bibliography{References}

\end{document}